\documentclass[twocolumn]{aastex62}

\hypersetup{linkcolor=red,citecolor=blue,filecolor=cyan,urlcolor=magenta}

\graphicspath{{./}{figures/}}

\shorttitle{The Mass Distribution of Superluminous Supernova Progenitors}
\shortauthors{Blanchard et al.}

\begin{document}

\title{The Pre-Explosion Mass Distribution of Hydrogen-Poor Superluminous Supernova Progenitors and New Evidence for a Mass-Spin Correlation}

\email{peter.blanchard@northwestern.edu}

\author{Peter K.~Blanchard}
\affil{Center for Interdisciplinary Exploration and Research in Astrophysics (CIERA), Northwestern University, 1800 Sherman Ave. 8th Floor, Evanston, IL 60201, USA}
\affil{Department of Physics and Astronomy, Northwestern University, 2145 Sheridan Rd., Evanston, IL 60208, USA}

\author{Edo Berger}
\affil{Harvard-Smithsonian Center for Astrophysics, 60 Garden St., Cambridge, MA 02138, USA}

\author{Matt Nicholl}
\affil{Birmingham Institute for Gravitational Wave Astronomy and School of Physics and Astronomy, University of Birmingham, Birmingham B15 2TT, UK}
\affil{Institute for Astronomy, University of Edinburgh, Royal Observatory, Blackford Hill, Edinburgh, EH9 3HJ, UK}

\author{V.~Ashley Villar}
\affil{Harvard-Smithsonian Center for Astrophysics, 60 Garden St., Cambridge, MA 02138, USA}

\begin{abstract}
Despite indications that superluminous supernovae (SLSNe) originate from massive progenitors, the lack of a uniformly analyzed statistical sample has so far prevented a detailed view of the progenitor mass distribution.  Here we present and analyze the pre-explosion mass distribution of hydrogen-poor SLSN progenitors as determined from uniformly modelled light curves of 62 events.  We construct the distribution by summing the ejecta mass posteriors of each event, using magnetar light curve models presented in our previous works (and using a nominal neutron star remnant mass).  The resulting distribution spans $3.6-40$ M$_{\odot}$, with a sharp decline at lower masses, and is best fit by a broken power law described by ${\rm d}N/{\rm dlog}M \propto M^{-0.41 \pm 0.06}$ at $3.6-8.6$ M$_{\odot}$ and $\propto M^{-1.26 \pm 0.06}$ at $8.6-40$ M$_{\odot}$.  We find that observational selection effects cannot account for the shape of the distribution.  Relative to Type Ib/c SNe, the SLSN mass distribution extends to much larger masses and has a different power-law shape, likely indicating that the formation of a magnetar allows more massive stars to explode as some of the rotational energy accelerates the ejecta. Comparing the SLSN distribution with predictions from single and binary star evolution models, we find that binary models for a metallicity of $Z\lesssim 1/3$ Z$_{\odot}$ are best able to reproduce its broad shape, in agreement with the preference of SLSNe for low metallicity environments.  Finally, we uncover a correlation between the pre-explosion mass and the magnetar initial spin period, where SLSNe with low masses have slower spins, a trend broadly consistent with the effects of angular momentum transport evident in models of rapidly-rotating carbon-oxygen stars.                      
\end{abstract}


\section{Introduction} 
\label{sec:intro}

The power source and progenitors of hydrogen-poor superluminous supernovae (Type I SLSNe; hereafter referred to as SLSNe) have remained a topic of intense debate since their discovery a decade ago \citep{Gal-Yam2009,Pastorello2010,Chomiuk2011,Quimby2011,Gal-Yam2012}, in part because of the order of magnitude spread in their peak luminosities and durations \citep{Nicholl2015,Lunnan2018,DeCia2018,Angus2019}. This has led to the suggestion of multiple energy sources \citep[e.g.,][]{Gal-Yam2012}, including abundant production of radioactive material and large ejecta masses from pair-instability explosions (PISNe; \citealt{HegerWoosley2002}), interaction of the SN ejecta with a dense circumstellar medium (CSM) that spans a range of mass loss rates and timescales \citep{ChevalierIrwin2011,Chatzopoulos2013}, and powering by a magnetar central engine \citep{KasenBildsten2010,Woosley2010}.  Indeed, for the latter case it has been shown that explosions with ejecta masses of $M_{\rm ej}\sim 2-20$ M$_\odot$ that produce magnetars with initial spins of $P\sim 1-10$ ms and magnetic fields of $B\sim 10^{13}-10^{15}$ G can explain the observed  range of peak luminosities and durations \citep{Inserra2013,Nicholl2014,Nicholl2017,Liu2017magLCs,Yu2017}.  In addition, late-time observations of SN\,2015bn indicate a light curve flattening consistent with the power-law spin down of a magnetar \citep{Nicholllate15bn}. Still, detailed light curve features such as post-peak undulations \citep{Nicholl2016a,Inserra2017,Blanchard2018} and early time bumps \citep{Leloudas2012,Nicholl2015a,Smith2016,NichollSmartt2016,Angus2019} remain to be understood.  

Unlike the other possible models, the magnetar interpretation is also supported by early-time ultraviolet/optical spectra \citep{Dessart2012,Mazzali2016,Nicholl2017gaia16apd} and by late-time optical nebular spectra, which show a remarkable similarity to broad-lined Type Ic SNe (SNe Ic-BL), in particular those associated with long gamma-ray bursts (LGRBs; \citealt{Milisavljevic2013,Nicholl2016b,Nicholl2019,Jerkstrand2016,Jerkstrand2017}).  In addition, late-time radio emission detected at the position of PTF10hgi, the first such detection for a SLSN, has been shown to be consistent with expectations from a magnetar wind nebula \citep{Eftekhari2019}.  On the other hand, there is currently no unambiguous case of a PISN origin for a SLSN. Similarly, while there is evidence for CSM interaction in a handful of events \citep{Yan2017}, including possibly the aforementioned light curve bumps, the inferred masses of this CSM are not sufficient to explain the peak luminosity and duration of SLSN light curves.  While some CSM interaction likely contributes to the diversity of SLSNe, the spectroscopic properties are difficult to reconcile with a model in which CSM interaction is the dominant power source\footnote{On the other hand, the recent SN\,2016iet, which differs from Type I SLSNe in its light curves and spectra, does provide strong evidence for powering by CSM interaction with a hydrogen-poor medium \citep{Gomez2019}.  The significant differences between SN\,2016iet and SLSNe is another line of evidence against dominant CSM interaction in SLSNe.} 

Following this recent progress in understanding the power source in SLSNe, a key outstanding question is what types of progenitors can produce them.  Given the lack of hydrogen and helium in the spectra of SLSNe, their progenitors are likely similar to those of other hydrogen-poor explosions such as Type Ib/c SNe (e.g., \citealt{Pastorello2010}) and LGRBs.  The low volumetric rate of SLSNe compared to the overall Type Ib/c SN rate \citep{Prajs2017}, and their preference for low metallicity dwarf host galaxies \citep{Chen2013,Lunnan2014,Leloudas2015,Perley2016,Schulze2018} points to a closer association with LGRB progenitors (another point in support of an engine-powered origin for SLSNe). In addition, a few SLSNe have exhibited direct links with LGRBs and SNe Ic-BL \citep{Greiner2015,Blanchard2019} and on average SLSNe have absorption velocities similar to those of SNe Ic-BL \citep{Liu2017}.  However, the details of this connection remain unclear.

Within the context of energy sources, progenitor properties, and the relation of SLSNe to other stripped SNe, it is essential to determine the distribution of pre-explosion progenitor masses, or nearly equivalently the ejecta masses.  Moreover, within the context of the magnetar model, it is equally critical to assess any correlations between the engine and explosion properties, which may shed light on the conditions for SLSN production. Such a study requires a uniform analysis of a sufficiently large sample, which is now emerging from large-scale surveys such as the Pan-STARRS1 Medium-Deep Survey \citep{Lunnan2018}, the Palomar Transient Factory \citep{DeCia2018}, the Dark Energy Survey \citep{Angus2019}, and the Zwicky Transient Facility \citep{Lunnan2019}.

Here we analyze the multi-band light curves of 62 SLSNe at $z\approx 0.06-1.6$ using the magnetar model (implemented in the light curve fitting code {\tt MOSFiT}; \citealt{Guillochon2018}) to determine the ejecta mass distribution as a proxy for the progenitor mass distribution at the time of explosion, as well as to explore underlying correlations between the explosion and engine properties.  Our study builds on the work and models of \citet{Nicholl2017}, which was focused on a smaller sample of 38 SLSNe, and \citet{Villar2018} which was focused on assessing SLSN science with the Large Synoptic Survey Telescope.  We furthermore utilize simulated SLSN light curves with the same models to assess and account for the impact of observational selection biases in the existing SLSN sample. 

The paper is structured as follows.  In \S\ref{sec:sample} we present the SLSN sample.  In \S\ref{sec:main} we determine and analyze the progenitor mass distribution and compare it to Type Ib/c SNe and to massive star (single and binary) evolutionary models.  In \S\ref{sec:corr} we present evidence for a correlation between the ejecta mass and magnetar initial spin.  In \S\ref{sec:disc} we discuss the implications for our understanding of SLSN progenitors, and we conclude in \S\ref{sec:concl}.

\section{SLSN Sample}
\label{sec:sample}

The sample of SLSNe in this paper consists of 62 events modeled uniformly using a Bayesian approach with {\tt MOSFiT} in \citet{Nicholl2017}, \citet{Villar2018}, \citet{Blanchard2018}, and \citet{Blanchard2019}.  We also include the SLSN PS16fgt, whose light curves and models will be presented in a forthcoming publication (Blanchard et al.~in preparation).  The details of the magnetar model, parameter priors, and Markov Chain Monte Carlo fitting procedure used for modeling the light curves are provided in \citet{Nicholl2017}.  In Table~\ref{tab:info} we provide the SLSN names, redshifts, original sources of the light curve data, and the references for the models.  Tables presenting the inferred parameters for each event are provided in the model reference papers.  The SLSNe in our sample span the redshift range $z\approx 0.06 - 1.6$, with most events at $z\lesssim 0.5$ (a result of a combination of the intrinsic volumetric rate of SLSNe and heterogenous survey depths and areal coverages).  In \S\ref{sec:dist} we assess any impact of redshift on the mass distribution.

\startlongtable
\begin{deluxetable*}{c|c|c|c}
\tablecolumns{4}
\tabcolsep0.05in\footnotesize
\tablewidth{3in}
\tablecaption{SLSN Sample   
\label{tab:info}}
\tablehead {
\colhead {SLSN}   &
\colhead {Redshift} &
\colhead {Data Reference} &
\colhead {Model Reference}
}   
\startdata
DES14X3taz & 0.608 & \citet{Smith2016} & \citet{Nicholl2017} \\ 
iPTF13ajg & 0.740 & \citet{Vreeswijk2014} & \citet{Nicholl2017} \\ 
iPTF13dcc & 0.431 & \citet{Vreeswijk2017} & \citet{Nicholl2017} \\
iPTF13ehe & 0.3434 & \citet{Yan2015} & \citet{Nicholl2017} \\ 
iPTF15esb & 0.224 & \citet{Yan2017} & \citet{Nicholl2017} \\ 
iPTF16bad & 0.2467 & \citet{Yan2017} & \citet{Nicholl2017} \\ 
LSQ12dlf & 0.255 & \citet{Nicholl2014} & \citet{Nicholl2017} \\ 
LSQ14bdq & 0.345 & \citet{Nicholl2015a} & \citet{Nicholl2017} \\ 
LSQ14mo & 0.253 & \citet{Chen2017} & \citet{Nicholl2017} \\ 
PS1-10ahf & 1.1 & \citet{McCrum2015} & \citet{Nicholl2017} \\ 
PS1-10awh & 0.908 & \citet{Chomiuk2011} & \citet{Nicholl2017} \\ 
PS1-10bzj & 0.650 & \citet{Lunnan2013} & \citet{Nicholl2017} \\ 
PS1-10ky & 0.956 & \citet{Chomiuk2011} & \citet{Nicholl2017} \\ 
PS1-10pm & 1.206 & \citet{McCrum2015} & \citet{Nicholl2017} \\ 
PS1-11afv & 1.407 & \citet{Lunnan2018} & \citet{Villar2018} \\ 
PS1-11aib & 0.997 & \citet{Lunnan2018} & \citet{Villar2018} \\ 
PS1-11ap & 0.524 & \citet{McCrum2014} & \citet{Nicholl2017} \\ 
PS1-11bam & 1.565 & \citet{Berger2012} & \citet{Nicholl2017} \\ 
PS1-11bdn & 0.738 & \citet{Lunnan2018} & \citet{Villar2018} \\ 
PS1-11tt & 1.283 & \citet{Lunnan2018} & \citet{Villar2018} \\ 
PS1-12bmy & 1.572 & \citet{Lunnan2018} & \citet{Villar2018} \\ 
PS1-12bqf & 0.522 & \citet{Lunnan2018} & \citet{Villar2018} \\ 
PS1-13gt & 0.884 & \citet{Lunnan2018} & \citet{Villar2018} \\ 
PS1-13or & 1.52 & \citet{Lunnan2018} & \citet{Villar2018} \\ 
PS1-14bj & 0.5215 & \citet{Lunnan2016} & \citet{Nicholl2017} \\ 
PS16aqv & 0.2025 & \citet{Blanchard2018} & \citet{Blanchard2018} \\ 
PS16fgt & 0.30 & Blanchard et al.~in prep. & Blanchard et al.~in prep. \\ 
PS17dbf & 0.13 & \citet{Blanchard2019} & \citet{Blanchard2019} \\ 
PTF09atu & 0.5015 & \citet{Quimby2011} & \citet{Nicholl2017} \\
         &        & \citet{DeCia2018} & \\
PTF09cnd & 0.2584 & \citet{Quimby2011} & \citet{Nicholl2017} \\ 
PTF10aagc & 0.206 & \citet{DeCia2018} & \citet{Villar2018} \\ 
PTF10bfz & 0.1701 & \citet{DeCia2018} & \citet{Villar2018} \\ 
PTF10nmn & 0.1237 & \citet{DeCia2018} & \citet{Villar2018} \\ 
PTF10uhf & 0.2882 & \citet{DeCia2018} & \citet{Villar2018} \\ 
PTF10vqv & 0.4518 & \citet{DeCia2018} & \citet{Villar2018} \\ 
PTF11hrq & 0.057 & \citet{DeCia2018} & \citet{Villar2018} \\ 
PTF12dam & 0.1073 & \citet{Nicholl2013} & \citet{Nicholl2017} \\
         &        & \citet{Chen2015} & \\
         &        & \citet{Vreeswijk2017} & \\
PTF12gty & 0.176 & \citet{DeCia2018} & \citet{Villar2018} \\ 
PTF12hni & 0.107 & \citet{DeCia2018} & \citet{Villar2018} \\ 
PTF12mxx & 0.3296 & \citet{DeCia2018} & \citet{Villar2018} \\ 
PTF13bjz & 0.271 & \citet{DeCia2018} & \citet{Villar2018} \\ 
PTF13cjq & 0.396 & \citet{DeCia2018} & \citet{Villar2018} \\ 
SCP-06F6 & 1.189 & \citet{Barbary2009} & \citet{Nicholl2017} \\ 
SN2005ap & 0.2832 & \citet{Quimby2007} & \citet{Nicholl2017} \\ 
SN2006oz & 0.376 & \citet{Leloudas2012} & \citet{Nicholl2017} \\ 
SN2007bi & 0.1279 & \citet{Gal-Yam2009} & \citet{Nicholl2017} \\ 
SN2009cb & 0.1867 & \citet{DeCia2018} & This paper \\ 
SN2009jh & 0.3499 & \citet{Quimby2011} & \citet{Nicholl2017} \\ 
         &        & \citet{DeCia2018} & \\
SN2010gx & 0.2297 & \citet{Pastorello2010} & \citet{Nicholl2017} \\ 
         &        & \citet{Quimby2011} & \\
SN2010hy & 0.1901 & \citet{DeCia2018} & This paper \\ 
SN2010md & 0.0987 & \citet{Inserra2013} & \citet{Nicholl2017} \\ 
         &        & \citet{DeCia2018} & \\ 
SN2011ke & 0.1428 & \citet{Inserra2013} & \citet{Nicholl2017} \\ 
SN2011kf & 0.245 & \citet{Inserra2013} & \citet{Nicholl2017} \\ 
SN2011kg & 0.1924 & \citet{Inserra2013} & \citet{Nicholl2017} \\ 
SN2012il & 0.175 & \citet{Inserra2013} & \citet{Nicholl2017} \\ 
SN2013dg & 0.265 & \citet{Nicholl2014} & \citet{Nicholl2017} \\ 
SN2013hy & 0.663 & \citet{Papadopoulos2015} & \citet{Nicholl2017} \\ 
SN2015bn & 0.1136 & \citet{Nicholl2016a} & \citet{Nicholl2017} \\ 
SN2016eay & 0.1013 & \citet{Nicholl2017gaia16apd} & \citet{Nicholl2017} \\ 
SNLS-06D4eu & 1.588 & \citet{Howell2013} & \citet{Nicholl2017} \\ 
SNLS-07D2bv & 1.50 & \citet{Howell2013} & \citet{Nicholl2017} \\ 
SSS120810 & 0.156 & \citet{Nicholl2014} & \citet{Nicholl2017} \\
\enddata
\end{deluxetable*}

\begin{figure*}
\centering
\includegraphics[scale=0.58]{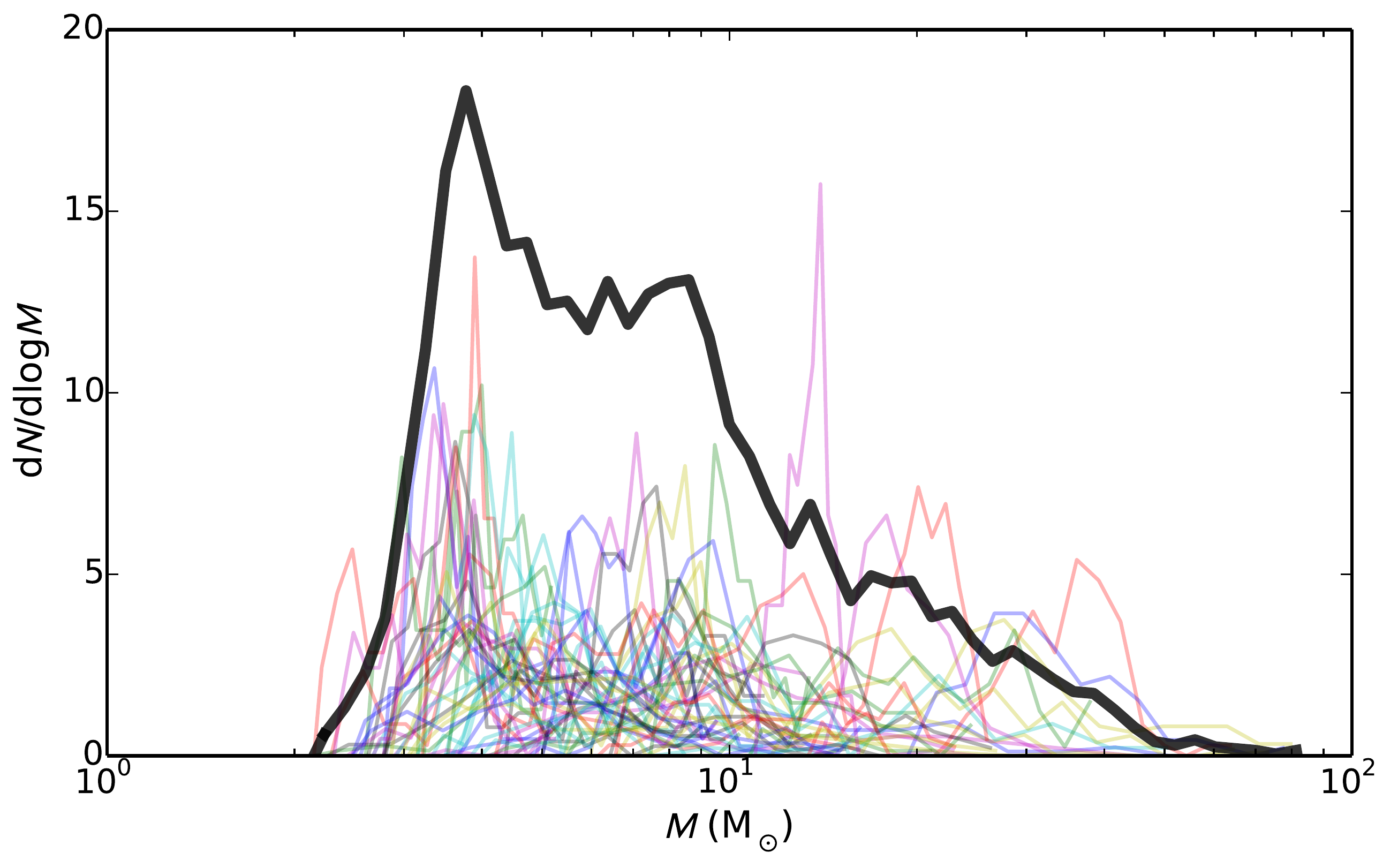}
\caption{Pre-explosion progenitor mass posterior distributions for each of the 62 SLSNe in our sample (i.e., the ejecta mass plus a nominal neutron star mass of 1.8 M$_{\odot}$) from which we construct the joint posterior of the full sample distribution (black line).  Most events have well-constrained (narrow) distributions.  By combining the posteriors of each SLSN, the final summed distribution correctly reflects the uncertainty for each event.}
\label{fig:posts}
\end{figure*}

\section{Progenitor Mass Distribution} 
\label{sec:main}

\subsection{Observed Characteristics}
\label{sec:dist}

The key outputs of the light curve models are the ejecta mass ($M_{\rm ej}$) and the neutron star's initial spin ($P$) and magnetic field strength ($B$).  To infer the progenitor masses at the time of explosion ($M$) we add a nominal neutron star remnant mass of $1.8$ M$_{\odot}$ to $M_{\rm ej}$.  While the range of masses of neutron stars formed after the core collapse of massive stars in general, and SLSNe in particular, is not precisely known, existing mass measurements of Galactic neutron stars suggest a range of $\sim 1.4-2.2$ M$_{\odot}$ \citep[e.g.~for a review see][]{NSreview}, making 1.8 M$_{\odot}$ a reasonable mass to use. This is consistent with the masses inferred by \citet{Nicholl2017}, who marginalized over neutron star mass as a free parameter in their fits.  We stress that varying the neutron star mass at the level of a few tenths of a solar mass does not affect our conclusions. 

In Figure~\ref{fig:posts} we plot the individual posterior distributions of $M$ for all 62 SLSNe in our sample.  The sample spans a broad range of masses, from about 3 to 40 M$_\odot$.  For some events the posteriors are well constrained, while for others a broader range of masses can be accommodated by the data.  In the following analysis we use the combined full posterior distribution of $M$, taking into account possible degeneracies with the engine parameters and other model nuisance parameters (e.g., opacities, dust extinction).  

We also plot in Figure~\ref{fig:posts} the full SLSN mass distribution, ${\rm d}N/{\rm dlog}M$, calculated by summing the mass posteriors for all 62 SLSNe.  The observed mass distribution exhibits several interesting features.  First, there is a sharp decline below about 3.6 M$_\odot$ (i.e., $M_{\rm ej}\approx 1.8$ M$_\odot$).  As we show below, this sharp decline is not due to an observational bias, and indeed Type Ib/c SN progenitors exhibit a similar turnover in their pre-explosion mass distribution.

\begin{figure}[!t]
    \centering
    \includegraphics[scale=0.41]{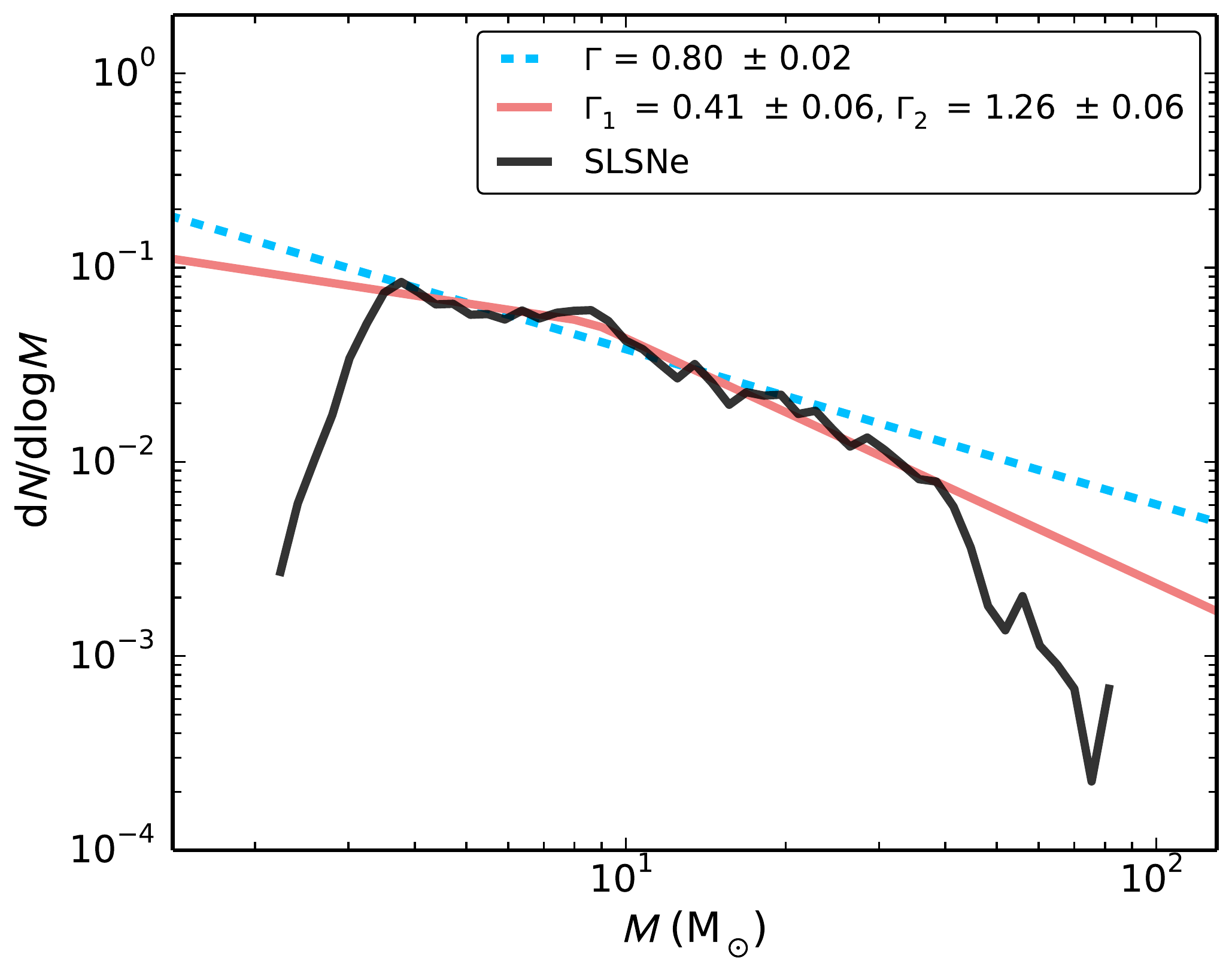}
    \caption{Pre-explosion progenitor mass distribution of SLSNe (black; from Figure~\ref{fig:posts}) compared to single (dashed blue line) and broken (solid red line) power-law fits.  The power-law index, $\Gamma$, defined in the equation ${\rm d}N/{\rm dlog}M \propto M^{-\Gamma}$, is given for the single power-law fit, as well as the corresponding indices, $\Gamma_{1}$ and $\Gamma_{2}$, for the broken power-law fit above and below the break mass of 8.6 M$_{\odot}$, respectively.}
    \label{fig:massdist}
\end{figure}

Second, at masses of $\approx 3.6$ to $\approx 40$ M$_\odot$ the distribution appears to follow a broken power law with a mild decline to about 9 M$_\odot$ and a sharper decline thereafter.  Fitting the distribution with such a model (Figure~\ref{fig:massdist}), where the two power-law indices and the location of the break are free parameters, we find ${\rm d}N/{\rm d log}M\propto M^{-0.41\pm 0.06}$ at $3.6-8.6$ M$_{\odot}$ and $\propto M^{-1.26\pm 0.06}$ at $8.6-40$ M$_{\odot}$.  A single power-law fit gives an index of $-0.80\pm 0.02$ (Figure~\ref{fig:massdist}), but it is inferior as indicated by assessing the goodness-of-fit using a testing procedure based on the Kolmogorov-Smirnov (KS) test statistic\footnote{Since the critical values associated with the standard application of the KS test are invalid when the comparison cumulative distribution function is estimated from the data, we perform a Monte Carlo procedure to estimate the appropriate null distribution of the KS statistic.  We generate $10^{4}$ random samples each drawn from a distribution given by the fitted single power law and calculate the KS statistic for each sample.  We then take the resulting distribution of KS statistics to be the null distribution of the KS statistic and calculate the $p$-value associated with the measured KS statistic of the SLSN mass distribution.}.  We find $p\lesssim 3\times10^{-3}$, indicating that we can reject at $>3\sigma$ significance the null hypothesis that the SLSN mass distribution is drawn from the fitted single power-law distribution.  Performing the same goodness-of-fit assessment for the broken power-law fit, we find $p\approx 0.16$ indicating the SLSN distribution is consistent with being drawn from a broken power-law distribution.  

Finally, we find a sharp decline at $\gtrsim 40$ M$_\odot$, but this is likely due to the finite sample size, with the higher masses representing the tails of the posterior distributions of a few individual events (Figure~\ref{fig:posts}), rather than an actual trend in the data.  We do note, however, that the drop-off occurs at the core mass where mass ejections due to the pulsational pair-instability (PPI) are expected to become important \citep{Woosley2017}.  The final explosions of stars that undergo PPI should therefore eject $\lesssim 40$ M$_{\odot}$, leading to a prediction for a drop-off in the pre-explosion mass distribution.  The robustness of this drop-off will be tested with a larger sample in the future. 

To assess for any redshift evolution in the SLSN mass distribution we divide the sample into low and high redshift bins using a range of redshift cuts (see Appendix).  While there is a hint that the lowest redshift events ($z\lesssim 0.25$) contain fewer high mass progenitors, the sample size at these redshifts is too small ($N=23$) to make a robust statement.  Using the median redshift of $z\approx 0.31$ to divide the sample, we find no significant change in the mass distribution from low to high redshift (i.e., dividing the sample by redshift is statistically consistent with dividing the sample randomly).    

\begin{figure}
\centering
\includegraphics[scale=0.35]{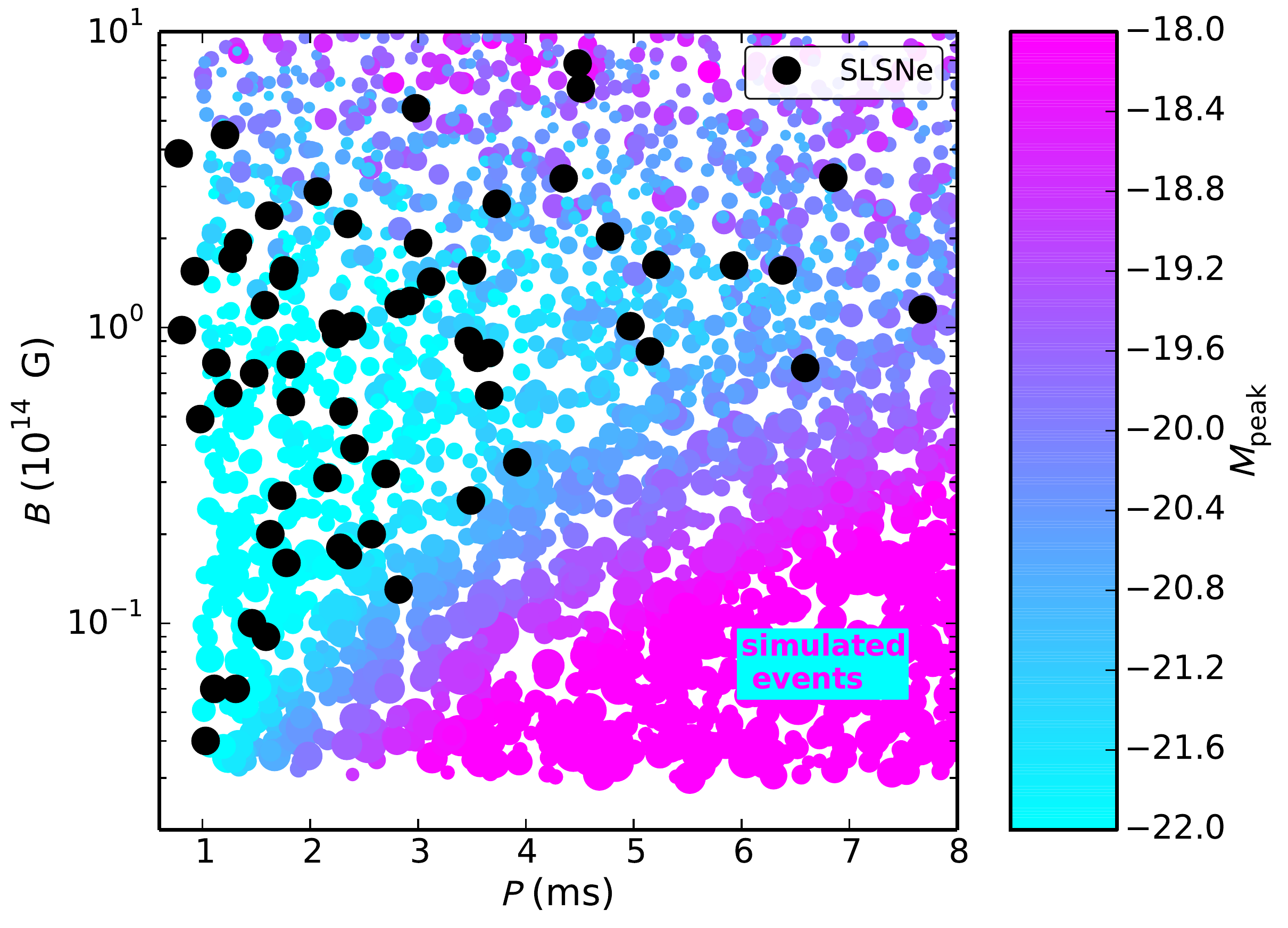}
\includegraphics[scale=0.42]{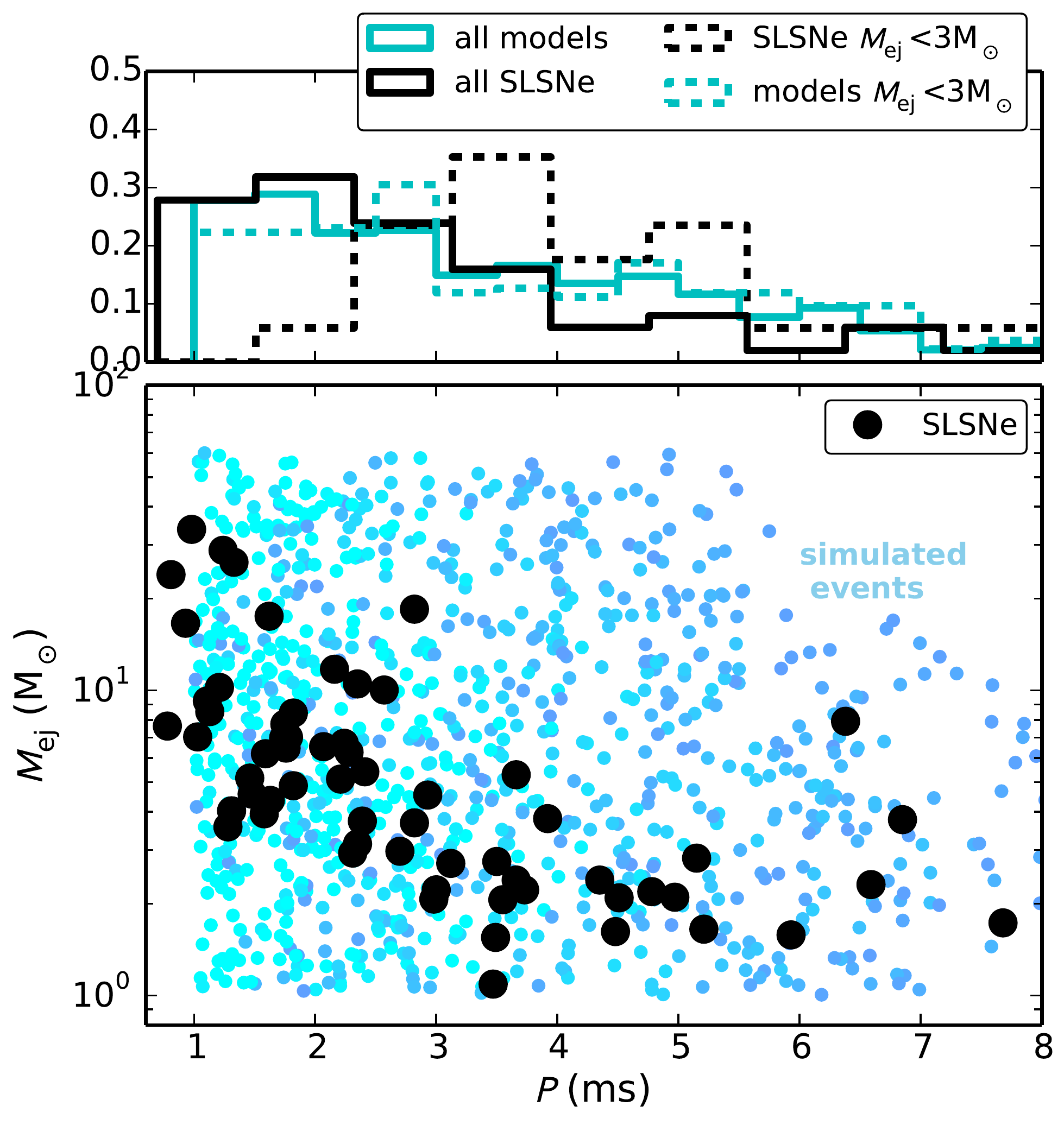}
\caption{Top: The $B-P$ parameter space showing the wedge filled by the observed sample (black points), which corresponds to $M_{\rm peak} \lesssim -20.5$.  Bottom: $M_{\rm ej} - P$ parameter space showing the parameter medians inferred for the observed sample of SLSNe (black points), as well as the parameters for the simulated models which satisfy $M_{\rm peak} < -20.5$ mag and $t_{\rm dur} > 30$ days (colored points; $t_{\rm dur}$ defined in the text).  We compare the projected spin histograms for the full observed (solid black) and simulated (solid cyan) samples to those for $M_{\rm ej}<3$ M$_{\odot}$ (dashed black and cyan).  While the observed spin distribution summed over all masses is overall similar to that from the models, at $M_{\rm ej}<3$ M$_{\odot}$, the distribution is shifted to slower spins compared to the models.}
\label{fig:mass_spin}
\end{figure}

\subsection{Selection Effects}
\label{sec:bias}

To understand how observational selection biases may affect the inferred progenitor mass distribution we simulate a grid of 3,000 magnetar-powered SLSN light curves, sampled from uniform distributions of $M_{\rm ej}$ ($1-60$ M$_{\odot}$), $P$ ($1-8$ ms), and $B$ ($0.03-10 \times 10^{14}$ G).  We furthermore assume a Gaussian ejecta velocity distribution with a mean and standard deviation based on observed SLSNe \citep{Liu2017}.  We fix the opacity and gamma-ray opacity to 0.16 cm$^{2}$ g$^{-1}$ and 0.04 cm$^{2}$ g$^{-1}$, respectively, motivated by typical values inferred for the observed SLSN sample \citep{Nicholl2017}.  To be consistent with our nominal neutron star mass above, we fix the neutron star mass for these models at 1.8 M$_{\odot}$.  For each simulated light curve we measure the peak $r$-band absolute magnitude, $M_{\rm peak}$, and the duration ($t_{\rm dur}$, defined as the timescale within 1 magnitude of peak).  While simulating the complex processes by which SLSNe are identified from heterogeneous optical time-domain surveys is beyond the scope of this paper, we test the overall effect of applying simple selection criteria to the model set.  

In particular, we select models with peak absolute $r$-band magnitude of $\lesssim -20.5$, chosen to match the distribution of the observed sample.  As shown in Figure~\ref{fig:mass_spin}, where we plot the model events and observed SLSNe in the $B-P$ parameter space, the observed sample indeed matches models with $M_{\rm peak}\lesssim -20.5$.  This selection effect leads to events that span a specific ``wedge'' in the $B-P$ parameter space, with a general bias against models with low $B$ values and slow spin, as well as a drop off at large values of $B\gtrsim 10^{15}$ G. This is because of the mis-match in spin-down and diffusion timescales at high $B$ and typical $E/M$ \citep{Nicholl2017}.

We further test the effect of SLSN light curve duration, in particular that fast evolving SLSNe may be missed either due to survey cadence restrictions or to an overall shape similarity to more common SN types (Ia, Ib/c).  To simulate this effect we select models with durations of $\gtrsim 30$ d, the approximate duration of a Type Ia SN.  In Figure~\ref{fig:mass_spin} we also plot the models which satisfy both the duration and luminosity cuts in $M_{\rm ej}-P$ parameter space, and compare to the observed events.

\begin{figure}
    \centering
    \includegraphics[scale=0.5]{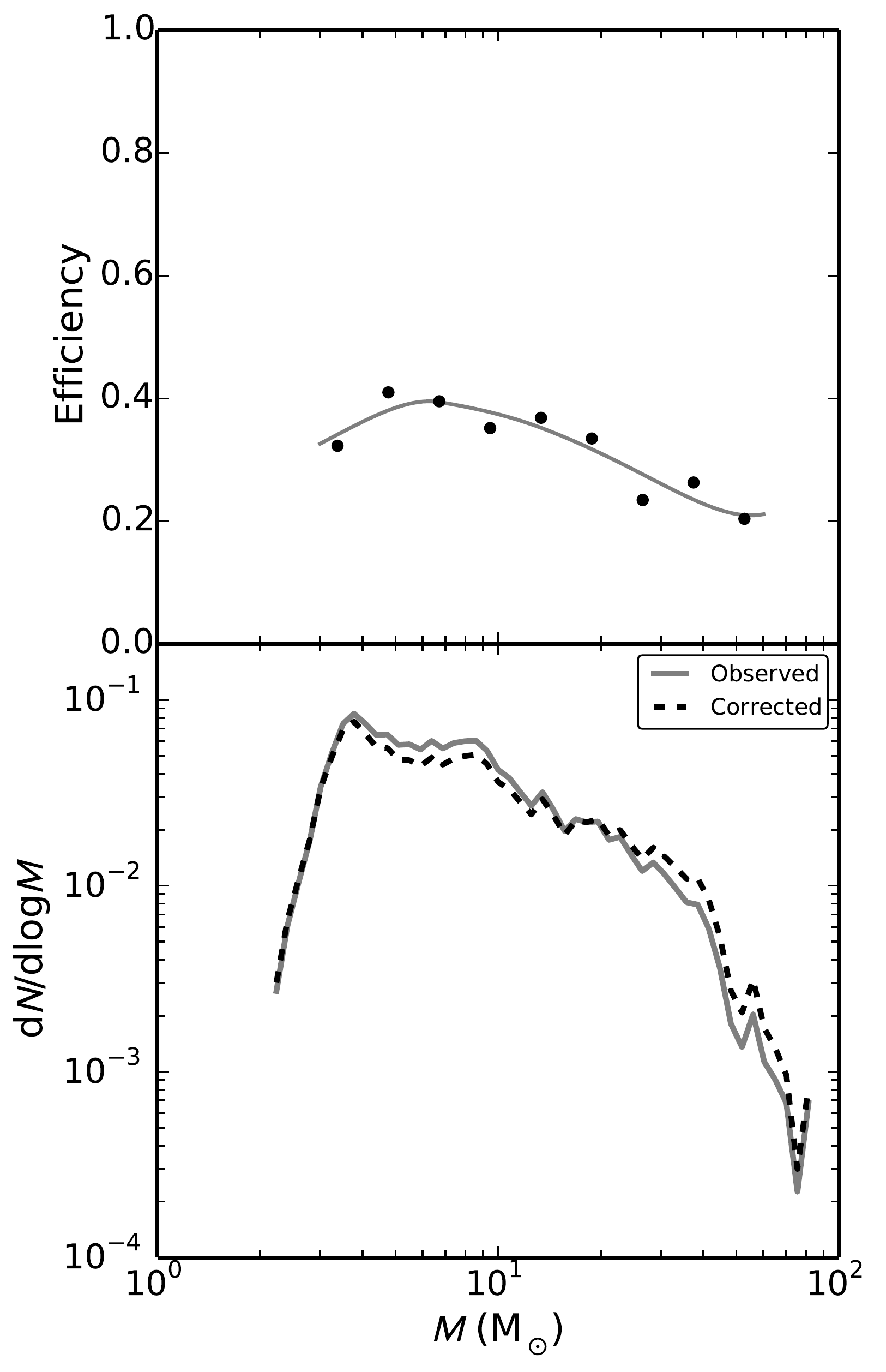}
    \caption{Top: Combined efficiency as a function of $M$ including both duration and luminosity effects calculated from our model light curves.  The slight decrease at low mass is a reflection of the duration cut, while the gradual decrease at high mass reflects the peak luminosity cut.  Bottom:  A comparison of the observed (grey solid line) and efficiency-corrected (black dashed line) SLSN pre-explosion progenitor mass distribution.}
    \label{fig:efficiency}
\end{figure}

With these simple, but observationally motivated, cuts on peak absolute magnitude and duration we assess the effect on the resulting distribution of $M$; namely, we calculate the SLSN detection efficiency as a function of $M$ by taking the ratio of the mass distribution of the model events with the cuts applied to the uniform input distribution.  As shown in Figure~\ref{fig:efficiency} we find that the efficiency slightly varies with mass.  There is a mild decrease at $\lesssim 6$ M$_\odot$ mainly due to resulting durations of $\lesssim 30$ d, and a gradual decrease in efficiency at higher masses due to an increasing fraction of events with peak magnitudes below $-20.5$ (apparent in the bottom panel of Figure~\ref{fig:mass_spin}).  However, the overall variation in efficiency across the full mass range is less than a factor of 2.

Applying this efficiency curve as a correction factor to the observed distribution of $M$ we find that the overall effect is minor, and primarily serves to slightly flatten the broken power-law shape of the uncorrected distribution (Figure~\ref{fig:efficiency}). Performing the same power-law fits as before, we find a single power-law index of $-0.72\pm 0.02$ (compared to $-0.80\pm 0.02$ for the uncorrected distribution), and a broken power-law fit with indices of $-0.54\pm 0.08$ and $-0.92\pm 0.05$ below and above the same break at 8.6 M$_{\odot}$ (also a free parameter), respectively.  Carrying out the same Monte Carlo KS goodness-of-fit procedure to assess the single power-law fit, we find $p\approx 8 \times 10^{-3}$, indicating that we can still rule out the possibility that the data are drawn from a single power law.   The broken power-law model has $p\approx 0.12$.

This indicates that the broken power-law shape of the distribution is statistically robust, and not a result of the basic selection effects considered here.  Therefore, the shape of the observed mass distribution can be taken to reflect the progenitor initial mass function (IMF) and subsequent stellar and binary evolution effects.  In the subsequent sections we compare the observed distribution to the implied pre-supernova mass distributions from stellar and binary evolution models to determine what type of progenitors are capable of reproducing the distribution.  

\subsection{Comparison to Type Ib/c SNe}
\label{sec:ibc}

\begin{figure}[!t]
    \centering
    \includegraphics[scale=0.41]{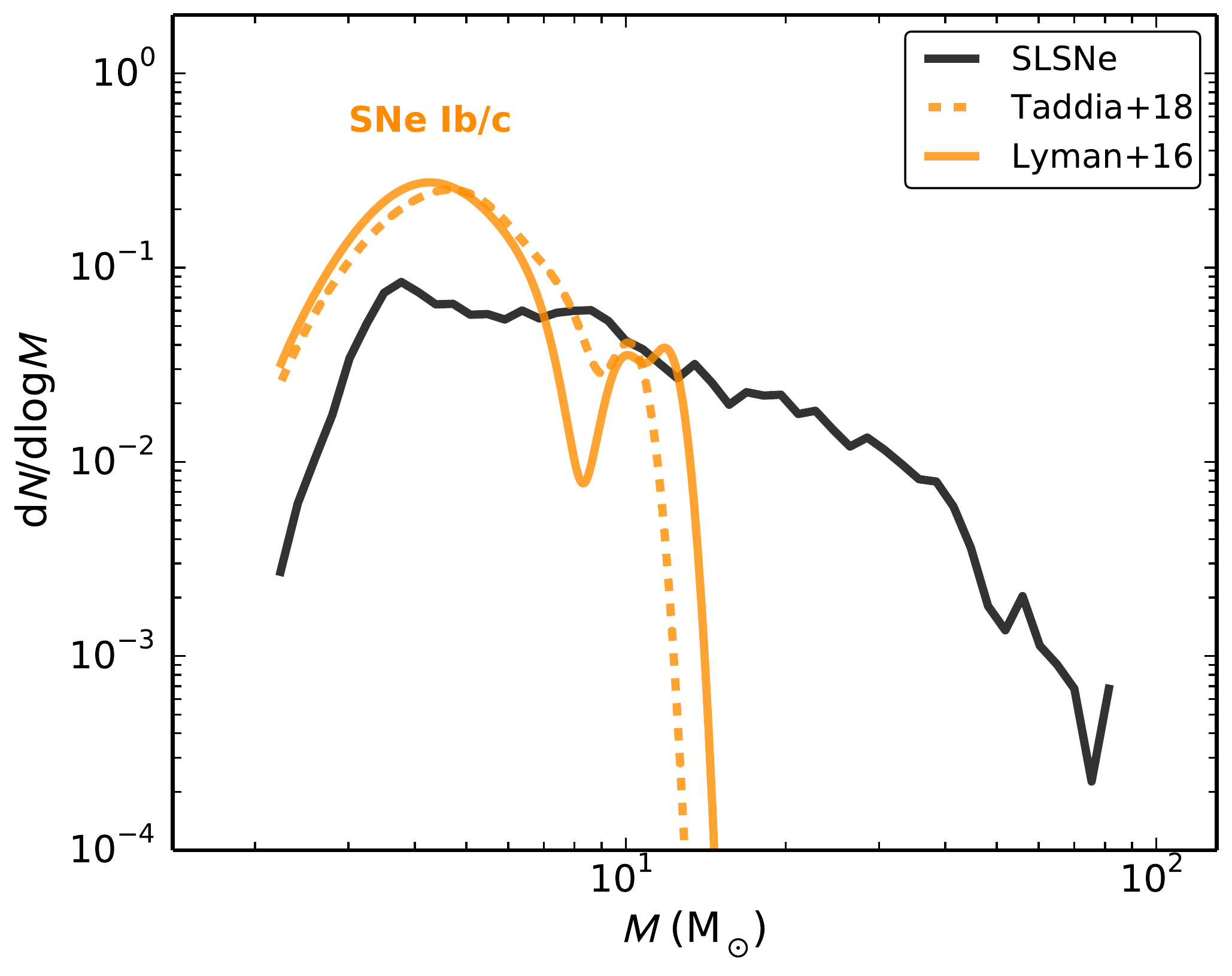}
    \caption{The pre-explosion mass distribution of SLSNe (black) compared to that for Type Ib/c SNe (solid orange: \citealt{Lyman2016}; dashed orange: \citealt{Taddia2018}).}
    \label{fig:compIbc}
\end{figure}

In Figure~\ref{fig:compIbc} we show a comparison of the SLSN pre-explosion mass distribution to the corresponding distribution for Type Ib/c SNe (SNe Ib/c).  We use the SN Ib/c samples from \citet{Lyman2016} and \citet{Taddia2018} who provide ejecta masses from the modeling of bolometric light curves.  Another recent sample from \citet{Prentice2019} is consistent with the results of these studies.  We add the same nominal neutron star mass of 1.8 M$_{\odot}$ to these values to infer the pre-explosion progenitor masses.  We create smooth distributions via kernel density estimation using a gaussian kernel with a bandwidth of 0.8 M$_{\odot}$, the typical uncertainty in the ejecta mass estimates.  The SLSN and SN Ib/c $M$ distributions exhibit both similarities and critical differences.  Both distributions exhibit a sharp decline at $\lesssim 3.5$ M$_\odot$ suggesting that in both cases the minimum progenitor masses are shaped by the same process, regardless of the ultimate fate of the explosion.

However, beyond the minimum cutoff the distributions differ.  The bulk of SNe Ib/c have pre-explosion progenitors in the range $M \approx 3-5$ M$_{\odot}$.  There are essentially no SN Ib/c progenitors with masses of $\gtrsim 10$ M$_\odot$, while SLSN progenitors extend to $\approx 40$ M$_\odot$.  In addition, between 3.5 and 10 M$_\odot$ the SN Ib/c distribution follows a much steeper (and single) power-law decline, ${\rm d}N/{\rm dlog}M \propto M^{-2.8}$, than the SLSN progenitors over this mass range (or in fact their entire mass range).

\begin{figure*}[!ht]
    \centering
    \includegraphics[scale=0.42]{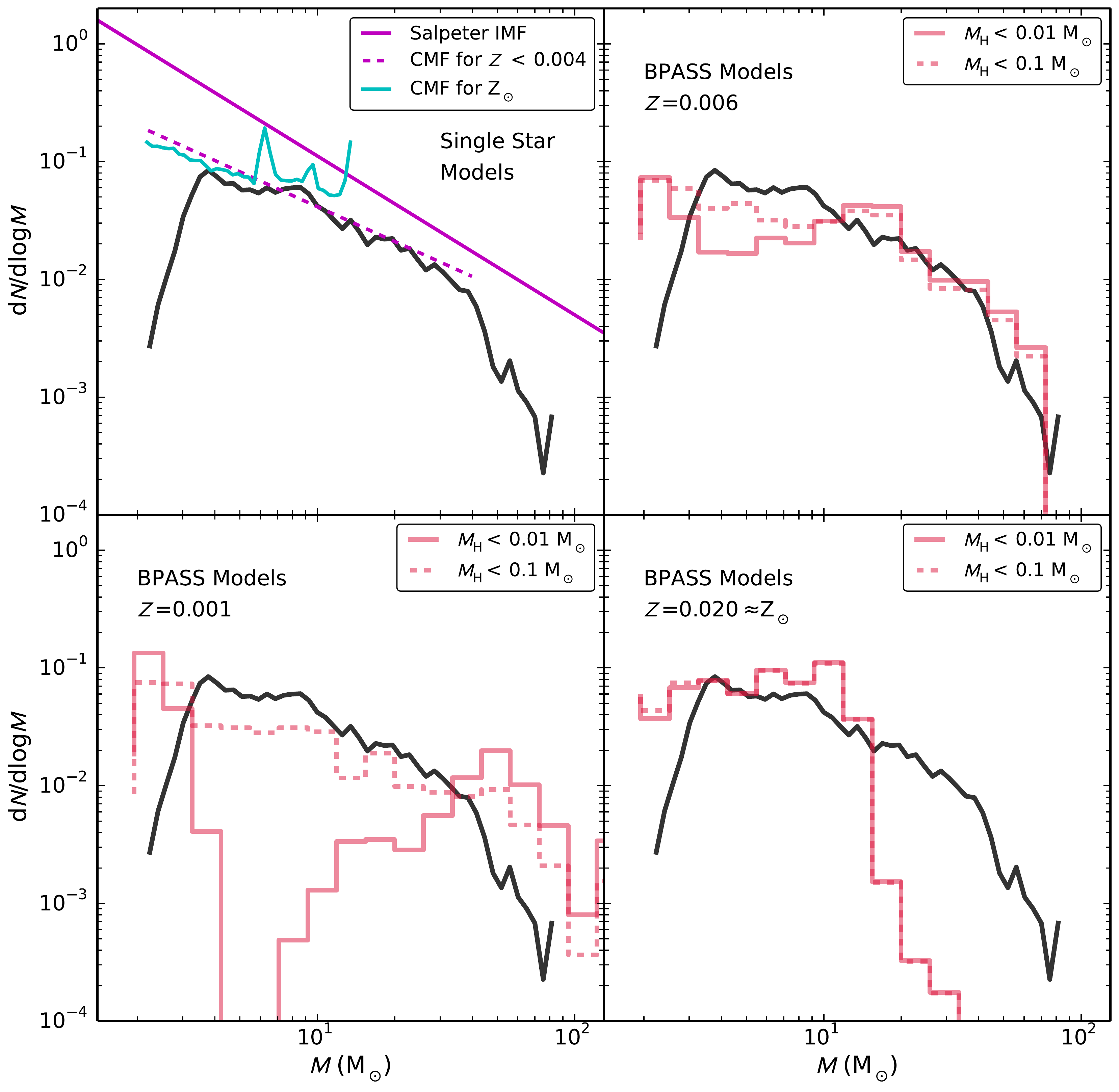}
    \caption{SLSN pre-explosion mass distribution (black line) compared to the distributions of CO core masses (CMFs) calculated using $M_{\rm ZAMS}-M_{\rm CO}$ relations from single star models (top left) at metallicities of $Z_{\odot}$ (solid cyan) and $Z<0.004$ (dashed magenta) and the distributions of final primary star masses from BPASS binary star models for metallicities of $Z=0.006$ (top right), $Z=0.001$ (bottom left), and $Z=0.020$ (bottom right).  We show the BPASS distributions for two cuts on the remaining hydrogen mass: $M_{\rm H} < 0.01$ M$_{\odot}$ and $M_{\rm H} < 0.1$ M$_{\odot}$.  For reference we also show the Salpeter IMF (solid magenta) in the top left panel.}
    \label{fig:modcomp}
\end{figure*}

\subsection{Comparison to Single Star Model Mass Functions}
\label{sec:single}

To understand the physics encoded by the shape of the pre-explosion progenitor mass distribution of SLSNe, we compare the distribution with the IMF, which for most variations follows the Salpeter power-law slope at high masses \citep{Salpeter1955}:
\begin{equation}
    \frac{{\rm d}N}{{\rm dlog}M} \propto M^{-1.35}.
\end{equation}
As shown in Figure~\ref{fig:modcomp}, the IMF matches well with the slope of the observed SLSN mass distribution at $\gtrsim 9$ M$_{\odot}$, but there is a clear deficit in the observed mass distribution at lower masses compared to the IMF expectation.

Since the IMF provides the distribution of zero-age main sequence (ZAMS) masses, while the pre-explosion mass distribution of SLSNe reflects significant mass loss and stripping, a more relevant comparison is to the core mass function (CMF).  Stellar evolution models indicate a dependence between ZAMS mass, core mass, and metallicity, which in turn influences mass loss.

We first compare the SLSN distribution with the expected CMF from stellar evolution models of solar metallicity stars, which experience significant mass loss due to line-driven winds.  We use models from \citet{Sukhbold2016}, which extend to ZAMS masses of 120 M$_{\odot}$.  The relation between core mass ($M_{\rm CO}$) and $M_{\rm ZAMS}$ for these models is a power law up to $M_{\rm ZAMS}\approx 40$ M$_{\odot}$ where $M_{\rm CO}$ peaks around 14 M$_{\odot}$.  Beyond $M_{\rm ZAMS}\approx 40$ M$_{\odot}$ the relation turns over and lower mass cores are produced due to increased mass loss.  In Figure~\ref{fig:modcomp} we show the resulting CMF for solar metallicity stars, which exhibits a relatively flat shape, with peaks caused by broad ranges of $M_{\rm ZAMS}$ that produce similar values of $M_{\rm CO}$, and an upper bound of about 14 M$_\odot$.  The CMF for solar metallicity stars is therefore not a good match to the SLSN distribution.  Solar metallicity models from \citet{Spera2015} are similarly unable to produce core masses of $\gtrsim 20$ $M_{\odot}$.  The fact that solar metallicity models cannot explain the SLSN mass distribution is consistent with the observed  preference of SLSNe in low metallicity ($Z\lesssim0.4$ Z$_{\odot}$) host galaxies \citep{Chen2013,Lunnan2014,Leloudas2015,Perley2016,Schulze2018}.

We next investigate low metallicity single star models.  Low metallicity stars generally lose less mass over their lifetimes and therefore yield CO core masses that increase monotonically with ZAMS mass.  Using the models of \citet{Sukhbold2018} with the lowest mass-loss rates (appropriate for low metallicity) we find that the relation between $M_{\rm CO}$ and $M_{\rm ZAMS}$ is a power law: 
\begin{equation}
    M_{\rm CO} \propto M_{\rm ZAMS}^{1.37}.
\end{equation}
A similar relation is obtained for the low metallicity models ($Z<0.004$) in \citet{Spera2015}, determined using a different stellar evolution code.  Combined with a Salpeter IMF this gives:
\begin{equation}
    \frac{{\rm d}N}{{\rm dlog}M}\propto M_{\rm CO}^{-0.99}.
    \label{eqn:co}
\end{equation}
In Figure~\ref{fig:modcomp} we plot the low metallicity CMF (truncated at the approximate core mass of $M \sim 40$ M$_{\odot}$ where the PPI is expected to become strong; \citealt{Woosley2017}) which exhibits much better agreement with the SLSN mass distribution.  Although in detail it does not strictly provide a broken power law shape that is preferred by the data. A key caveat of this comparison, however, is that the low metallicity models that yield this CMF predict that the stars will retain much of their hydrogen envelope at the time of explosion, whereas SLSN progenitors have been stripped of their envelopes.

In summary, for single star models, only low metallicity stars can generally account for the full range of pre-explosion (CO core) masses observed for SLSNe, as well as roughly the overall shape of the progenitor mass distribution.  However, the single star low metallicity models retain a significant hydrogen envelope at the time of explosion, making them incompatible with SLSN progenitors.  Recently \citet{Woosley2019} explored the evolution of hydrogen-poor stars using models of helium stars that have already lost their hydrogen envelopes.  Subsequent mass-loss was modeled assuming solar metallicity.  The resulting relation between the pre-supernova mass and $M_{\rm ZAMS}$ is similar in form to Equation~\ref{eqn:co} (their Equations 13 and 14), but with masses that are about half of $M_{\rm CO}$ in single star models with a hydrogen envelope.  Thus, in the \citet{Woosley2019} models explaining pre-explosion SLSN progenitors with $\gtrsim 20$ M$_{\odot}$ requires initial helium stars with $M_{\rm He}\gtrsim40$ M$_{\odot}$ (corresponding to ZAMS masses $\gtrsim 80$ M$_{\odot}$).  Importantly, this relation is also unable to account for the broken power law shape of the observed SLSN mass distribution.  At low metallicity, and therefore lower mass-loss rates, less massive helium stars are needed to explain such massive pre-explosion cores.  A $Z=0$ helium star with an initial mass of 30 M$_{\odot}$ modelled in \citet{Woosley2017} yielded a CO core mass of $\approx25$ M$_{\odot}$. 

A key ingredient of the magnetar model for SLSNe is rapid rotation.  None of the models discussed above include the effects of rotation on the late evolution.  Recently \citet{AguileraDena2018} explored the evolution of rapidly rotating stars, including models with enhanced rotational mixing that experience chemically homogeneous evolution (CHE) and leave behind relatively bare CO cores, potentially relevant for SLSNe. We find that the relation between $M_{\rm CO}$ and $M_{\rm ZAMS}$ implied by their models is also nearly linear, although these models produce larger CO core masses for a given $M_{\rm ZAMS}$ than the non-rotating models, which is in agreement with other models of rapidly rotating stars (e.g., \citealt{Yoon2006,ChatzopoulosWheeler2012,Woosley2017}).  This means that less massive initial stars are required to produce pre-explosion progenitors in the mass range observed for SLSNe.  For example, their model with an initial mass of 39 M$_{\odot}$ yields a final mass of $\approx 22$ M$_{\odot}$, whereas non-rotating models required $\approx80$ M$_{\odot}$ to produce such massive cores.  While these models are promising for SLSNe, the mass ``resolution'' of the implied relation from these models (they present models at only 6 specific masses) is not sufficient for a robust comparison with the detailed shape of the observed distribution.

\subsection{Comparison to Binary Star Models}
\label{sec:binary}

Another proposed mechanism for producing a bare CO core while maintaining rapid rotation is through interaction in a binary system.  It has long been suspected that binary interactions may be the dominant mechanism by which normal stripped SNe lose their hydrogen envelopes \citep[e.g.,][]{Podsiadlowski1992}, and the ejecta mass distributions of normal SNe Ib/c have been argued to be consistent with predictions from binary star models \citep{Lyman2016}.  In Figure~\ref{fig:modcomp} we compare our observed SLSN progenitor mass distribution with mass distributions from models of binary stars.  We use models of primary stars from the Binary Population and Spectral Synthesis (BPASS) suite of binary evolution models \citep[Version 2.2;][]{Eldridge2017,StanwayEldridge2018}.  In particular we use the ``z001'', ``z006'', and ``z020'' models, which are calculated for metallicities of $Z = 0.001$ ($\approx 1/20$ Z$_{\odot}$), $Z = 0.006$ ($\approx 1/3$ Z$_{\odot}$), and $Z = 0.020$ ($\approx$ Z$_{\odot}$), respectively.  We select those models that have remaining hydrogen masses of $M_{\rm H}<10^{-2}$ M$_{\odot}$ to restrict to models that produce stripped-envelope progenitors.  In Figure~\ref{fig:modcomp} we show the distributions of final primary star masses, weighted by an input Salpeter IMF, at each metallicity.  We also show the equivalent distributions for a relaxed cut on remaining hydrogen of $M_{\rm H}<10^{-1}$ M$_{\odot}$ to understand its effect on the distributions.

We find that the resulting distribution for $Z=0.006$ matches the observed SLSN mass distribution relatively well at $\gtrsim 10$ M$_{\odot}$, and also exhibits a flattening at lower masses as seen in the data, but with a significant deficit in expected numbers compared to the data that depends on the cut on $M_{\rm H}$.  The BPASS model distribution of stripped primary stars extends to the highest progenitor masses we find for SLSNe.  At a lower metallicity of $Z=0.001$ we find a gap between two populations of stripped ($M_{\rm H} < 10^{-2}$ M$_{\odot}$) stars at $M\sim2-3$ M$_{\odot}$ and $M\gtrsim10$ M$_{\odot}$, in stark contrast to the observed SLSN distribution, likely an indication that the low mass loss rates at these metallicities prevent the formation of stripped progenitors in the range $M\sim3-10$ M$_{\odot}$.  The solar metallicity models yield pre-explosion progenitors which span a similar range as the solar metallicity single star models, including a similarly steep drop-off at $M\gtrsim10$ M$_{\odot}$.  Therefore, among the three metallicities considered, the $Z=0.006$ models (representative of measured SLSN host galaxy metallicities) are best able to reproduce the observed SLSN distribution.  While the $Z=0.006$ distribution of stripped primary stars extends to the highest progenitor masses we find for SLSNe, the ability of the resulting explosions to actually eject most of the mass depends on the SN kinetic energy. 

\begin{figure}[!t]
    \centering
    \includegraphics[scale=0.405]{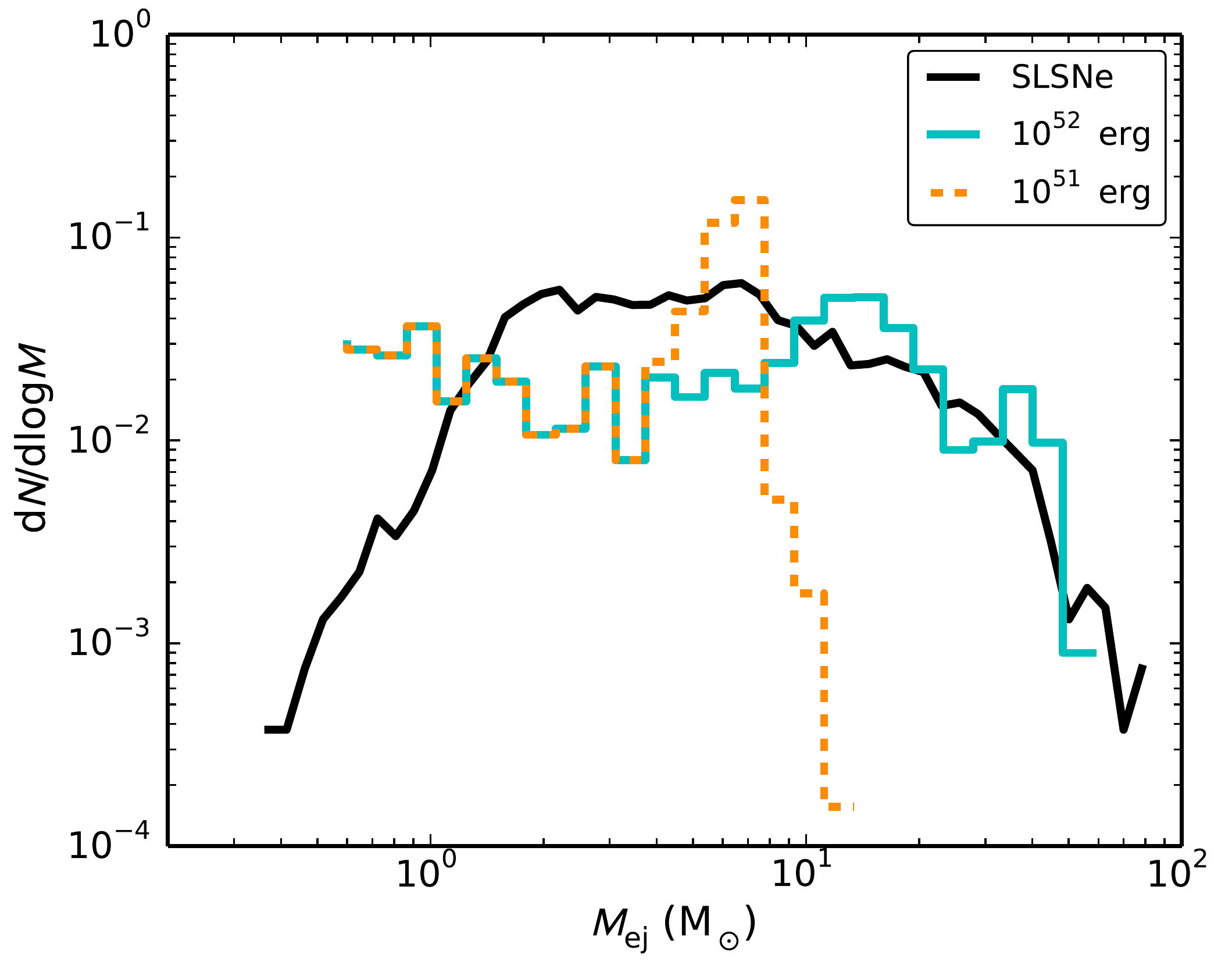}
    \caption{Ejecta mass distribution (i.e., calculated without adding the nominal remnant mass to each event) of our SLSN sample (solid black) compared to the ejecta masses from stripped primary star progenitors from BPASS models at metallicity $Z = 0.006$.  The model ejecta masses are calculated for both a kinetic energy of 10$^{51}$ erg (dashed orange) and 10$^{52}$ erg (solid cyan). 
    }
    \label{fig:bpass_mejs}
\end{figure}

To explore this effect, in Figure~\ref{fig:bpass_mejs} we directly compare the distribution of $M_{\rm ej}$ for SLSNe with BPASS models at $Z=0.006$ calculated for SN kinetic energies of $10^{51}$ and $10^{52}$ erg.  We find that the ejecta masses in the $10^{51}$ erg models rapidly drop off at $\approx 8$ M$_\odot$.  However, the $10^{52}$ erg models lead to the large ejecta masses seen in the observed SLSN distribution.  This therefore indicates that if SLSN progenitors evolve via a binary channel similar to the BPASS models, then a significant fraction of the events require much larger explosion energies than for normal SNe.  The magnetar models generally predict large kinetic energies with a median near $5\times 10^{51}$ erg, and many events with energies reaching $\sim 10^{52}$ erg.  This is due to conversion of the magnetar spin-down energy to ejecta kinetic energy.

\section{Mass-Spin Correlation}
\label{sec:corr}

In \S\ref{sec:bias} we explored the impact of observational selection biases on the observed pre-explosion mass distribution by analyzing the luminosity and duration of a grid of simulated magnetar light curves in relation to the observed properties of SLSNe.  While these selection effects have minor impact on the mass distribution, they help us to uncover an interesting trend between $M_{\rm ej}$ and $P$ in the observed sample that is not present in the simulated light curves.  In Figure~\ref{fig:mass_spin} we plot the observed and simulated events in the parameter space of $M_{\rm ej}$ and $P$, after removing simulated events with a low peak brightness of $\gtrsim -20.5$ mag and a short duration of $\lesssim 30$ d.  In the observed sample we find an overall correlation between $M_{\rm ej}$ and $P$, such that events with rapid spin tend to have higher ejecta masses than those with slower initial spin.  In particular, we find a dearth of SLSNe with $M_{\rm ej}\lesssim 3$ M$_{\odot}$ and $P\lesssim 3$ ms.  Conversely, we find that nearly all events with $M_{\rm ej}\gtrsim 8$ M$_{\odot}$ have $P\lesssim 3$ ms.

A comparison to the simulated events, after accounting for the main observational selection effects, indicates that the paucity of observed SLSNe with large $M_{\rm ej}$ and slow $P$ is at least partly due to an observational selection effect since these events tend to have systematically lower peak luminosities (Figure~\ref{fig:mass_spin}).  On the other hand, the lack of observed SLSNe with low $M_{\rm ej}$ and fast $P$ cannot be explained as an observational bias, with a significant fraction of the simulated events residing in this corner of parameter space.  

\begin{figure}[t!]
\centering
\includegraphics[scale=0.41]{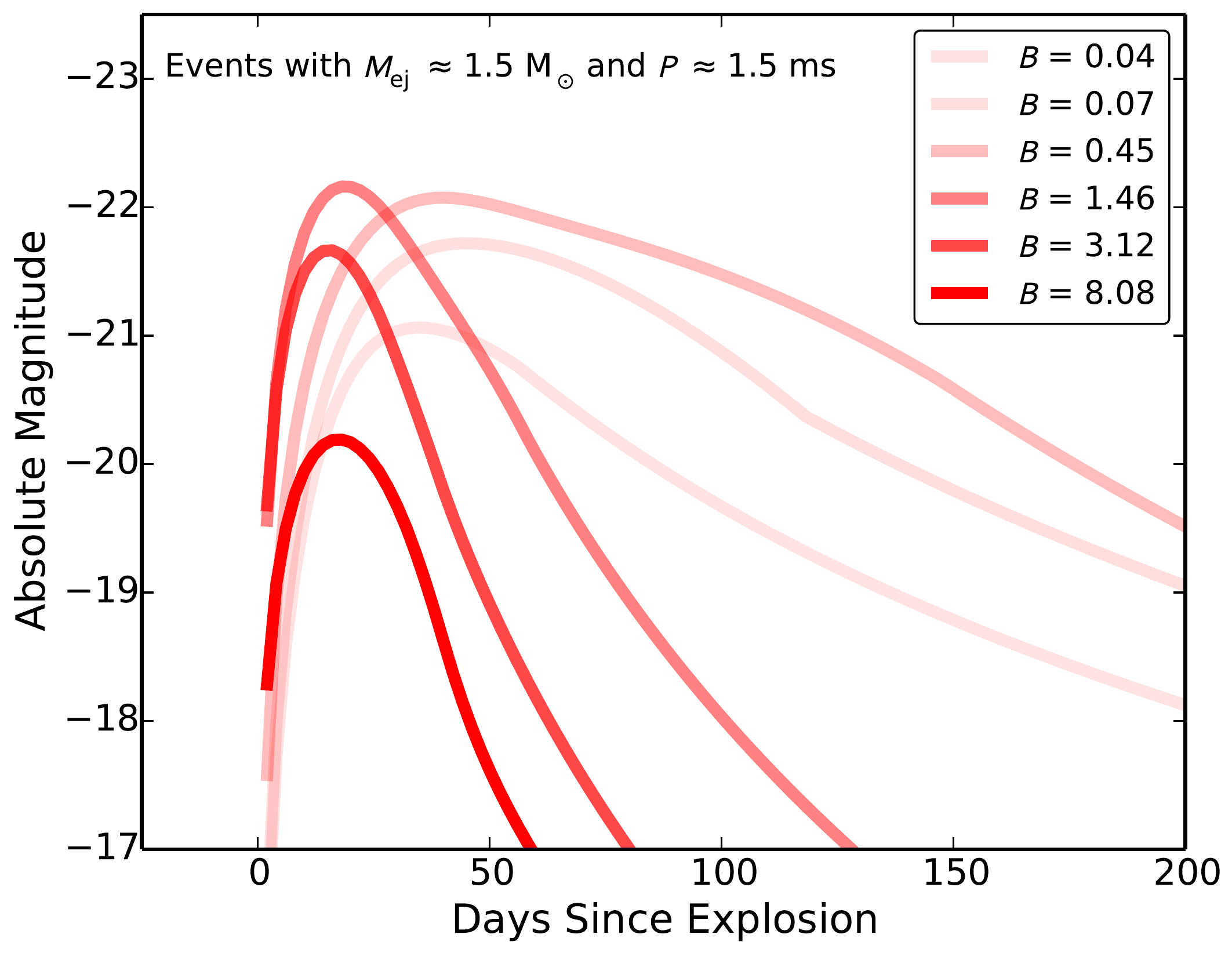}
\caption{Model $r$-band light curves from the region of parameter space with low $M_{\rm ej}$ and low $P$, showing the effect of magnetic field strength on duration ($B$ is given in units of 10$^{14}$ G).  While some events in this region evolve rapidly and may be missed by surveys, most are of sufficient duration to be captured.  Therefore the lack of observed events in this region of parameter space is not simply the result of an observational bias.
}
\label{fig:magLCs}
\end{figure}

In Figure~\ref{fig:magLCs} we show sample model light curves for events with $M_{\rm ej}\approx1.5$ M$_{\odot}$ and $P\approx1.5$ ms for a wide range of magnetic field strengths.  Most of these events have high peak luminosities and sufficiently long durations to be captured in optical sky surveys.  We therefore conclude that the lack of observed SLSNe with $M_{\rm ej}\lesssim 3$ M$_{\odot}$ and $P\lesssim 3$ ms is a real physical effect, and not an observational bias.  

\begin{figure}[t!]
\centering
\includegraphics[scale=0.41]{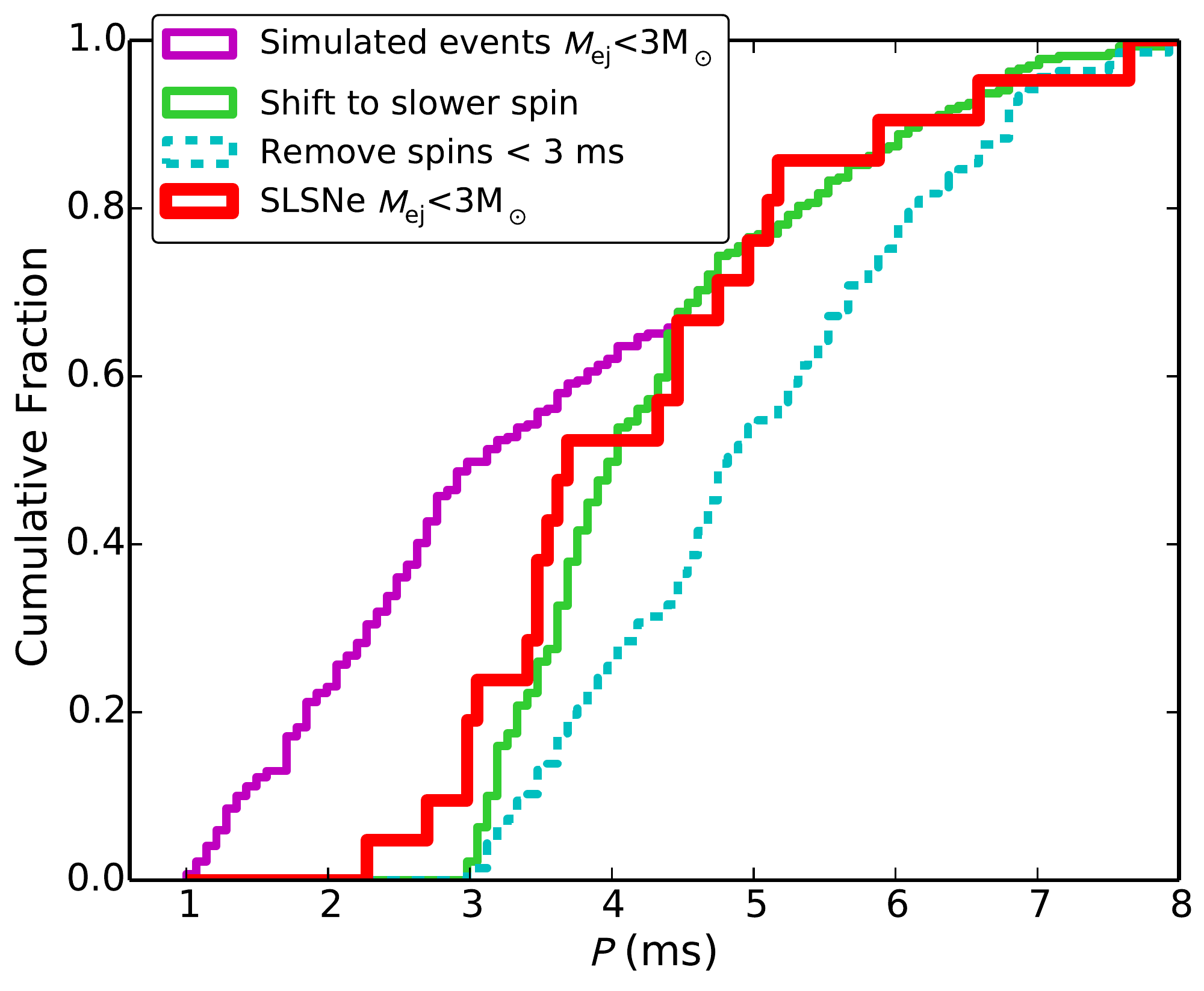}
\caption{Cumulative $P$ distributions for the models (with duration and luminosity cuts applied) and observed sample at $M_{\rm ej} < 3$ M$_{\odot}$, showing the significant difference at low $P$.  We also show distributions of the models after shifting models with $P < 3$ ms to slower spins or removing them completely from the model set.}
\label{fig:spindists}
\end{figure}

We simulate two possible effects to test whether they can account for this: (i) removing low ejecta mass ($\lesssim3$ M$_{\odot}$) events with $P<3$ ms from the simulated set, and (ii) systematically shifting such model events to slower spins (implemented by randomly shifting their spins to the range $P = 3-4.5$ ms).  These effects are designed to explore the possibilities that low mass progenitors with rapid spin are either not produced, or that they are spun down by $\approx1-2$ ms.  In Figure \ref{fig:spindists} we compare the cumulative spin distributions for the observed and simulated SLSNe with $M_{\rm ej}\lesssim 3$ M$_{\odot}$, which are significantly different as already hinted at above.  As shown in the Figure, the resulting simulated distribution after shifting model events to slower spins provides a better match to the observed sample than the distribution after completely removing them from the sample.   

This simple exercise lends support to the idea that the correlation between $M_{\rm ej}$ and $P$ is not simply the result of events being missed by surveys.  While we lack a detailed physical model that might explain this correlation, the effect of shifting events to slower spins could potentially be physically motivated by a process in which angular momentum is removed due to mass loss.  The lowest mass SLSN pre-explosion progenitors may be stars which have undergone enhanced mass loss relative to their more massive counterparts, leading to slower remnant spins.

\section{Discussion and Implications} 
\label{sec:disc}

In \S\ref{sec:dist} we found that the pre-explosion progenitor mass distribution of SLSNe exhibits a steep decline at $M\lesssim 3.6$ M$_{\odot}$ and a broken power-law shape with indices of $-0.41$ at $3.6$\,M$_{\odot}$ $<M<8.6$\,M$_{\odot}$ and $-1.26$ at $8.6$\,M$_{\odot}$ $<M<40$\,M$_{\odot}$.  While the broad shape is similar to the expected shape of the $M_{\rm CO}$ distribution (assuming an initial population following the Salpeter IMF) implied by models with fine mass grids of non-rotating low-metallicity single massive stars, as well as models of non-rotating helium stars, none of the non-rotating single star stellar models we explored provide a clear prediction that matches the detailed shape of the observed distribution.  This is consistent with the expected importance of rapid rotation.  

Confirming initial conclusions from a smaller sample \citep{Nicholl2015}, our comparison with SNe Ib/c shows that while many SLSNe result from progenitors of similar masses as SNe Ib/c, a significant fraction of SLSNe originate from progenitors with $M\gtrsim10$ M$_{\odot}$ and are therefore much more massive than a typical SN Ib/c progenitor.  Our larger sample provides further insight.  Namely, the power law shape of the distributions where they do overlap is different, with SNe Ib/c following a steeper slope than SLSNe.  These observations are a robust indication that the progenitors of SNe Ib/c and SLSNe exhibit different properties which impact the mass of stars that can explode.  These properties are likely related to the different evolutionary pathways (e.g., due to rapid rotation at low metallicity) which enable the production of a magnetar capable of enhancing the radiative output.  

The BPASS models suggest that achieving a SN with $M_{\rm ej} \gtrsim 8-10$ M$_{\odot}$ requires a large kinetic energy, which may only be possible if a magnetar is formed.  This comes from a binding energy argument for the primary star, and is thus not strictly a consequence of its prior binary evolution.  Otherwise, assuming a canonical explosion energy of $\sim10^{51}$ erg, core-collapses of such massive stars do not have enough energy to unbind most of their mass, leading to the formation of black holes \citep{Heger2003} and possibly fallback accretion powered SNe \citep{DexterKasen2013}.  The presence of a magnetar in SLSNe is therefore consistent with the broadening of the mass distribution relative to SNe Ib/c as the extra kinetic energy provided by the spin-down power allows for a more massive star to explode.  To summarize, the broad mass distribution that we observe for SLSNe is evidence that rotational energy is capable of not only enhancing the radiative output of a SN, but also the mass range of stars which can actually explode.  This is in contrast to the CSM interaction model for SLSNe, in which only the efficiency with which the available energy is converted to radiation is increased.   

Within the stripped-envelope SN class, SLSNe appear to be most similar to SNe Ic-BL in their mean spectroscopic \citep{Liu2017,Nicholl2019} and host environment properties \citep{Lunnan2014}.  In addition, several direct links have been found, including the SLSN SN\,2017dwh which evolved from a blue SLSN-like spectrum (with unusually strong absorption from \ion{Co}{2}) to appear nearly identical to SNe Ic-BL \citep{Blanchard2019}.  Also, the luminous ($M\approx-20$ mag) SN\,2011kl was found to be associated with the ultra-long GRB\,111209A \citep{Greiner2015}.  While SN\,2011kl was bluer and relatively featureless compared to previous SNe Ic-BL known to be associated with LGRBs, its discovery suggested a possible continuum of behavior from SNe Ic-BL to SLSNe with the implication being that these transients are linked by a common central engine mechanism \citep{Kann2019}.

It is therefore of particular interest to compare the pre-explosion progenitor masses of SLSNe with those of SNe Ic-BL.  While a similarly detailed distribution for SNe Ic-BL does not yet exist, light curve and spectroscopic analyses suggest on average masses higher than normal Type Ic SNe \citep{Drout2011,Lyman2016,Prentice2019}.  Estimates based on spectroscopic models for SNe Ic-BL associated with LGRBs are generally at the high mass end (e.g., $\approx$ 10 M$_{\odot}$ for SN\,1998bw, SN\,2003lw, and SN\,2003dh; \citealt{Iwamoto1998,Mazzali2013}), similar to many SLSNe.  This combined with the insight from events like SN\,2017dwh, nebular spectra, and environmental comparisons, suggests that the progenitor pathways which produce SLSNe are more similar to those that produce LGRBs and their associated SNe Ic-BL than those that produce normal SNe Ib/c.    

\subsection{Shape of the Mass Distribution as a Clue for Progenitor Formation}

Evolutionary models of non-rotating low metallicity stars generically predict a power-law relationship between $M_{\rm CO}$ and $M_{\rm ZAMS}$, which leads to a power-law distribution of $M_{\rm CO}$.  Interestingly, our single power-law fit to the SLSN pre-explosion mass distribution gives a slope similar to the expectation for non-rotating low-metallicity stars.  However, we find that a broken power law is a better description of the SLSN distribution, indicating a deviation from the simple $M_{\rm CO}$ distribution.  This suggests that rotation and mass-loss processes significantly affect the distribution of pre-SN CO core masses.    

In the context of a magnetar engine, rapidly rotating CO cores appear to be essential.  Mass-loss and spin-up processes resulting from binary interactions \citep{deMink2013} as well as rotationally-induced mixing associated with CHE \citep{YoonLanger2005,WoosleyHeger2006} have been proposed to explain how mostly bare rapidly rotating CO cores can be produced.  In addition, some progenitors may undergo additional mass loss which removes mass from the CO cores.  Recent models of such stars, produced via CHE, suggest that spin up from core contraction induces mass loss \citep{AguileraDena2018}.  These authors suggest that possible signatures of CSM interaction seen in several SLSNe (e.g., light curve bumps; \citealt{Nicholl2016a,Inserra2017,Yan2017,Blanchard2018}) is evidence of this mass-loss process and that it may occur regardless of how rapidly rotating CO stars are formed.
Mass loss is therefore a key factor which influences the shape of the pre-explosion SLSN progenitor mass distribution.

Additional effects due to binary interaction are likely given that observations suggest about 70\% of O stars experience mass transfer with a binary companion \citep{Sana2012}.  \citet{Woosley2019} approximated the effects of binary interaction by studying the evolution of helium stars which have already lost their hydrogen envelope.  However, we find that these models still predict a smooth distribution of $M_{\rm CO}$.  Moreover, this study does not account for the effects of rotation and only incorporates further mass loss due to stellar winds at solar metallicity.  We also compare our distribution to BPASS binary models which simultaneously take into account the effects of stellar winds, mixing due to rotation, Roche lobe overflow, and common envelope evolution.  We find the mass distribution of stripped primary stars at $Z\approx 1/3$ Z$_{\odot}$ is relatively flat (in ${\rm dlog}M$) at $\lesssim 20$ M$_{\odot}$, similar to the break structure we find for SLSNe though with a higher break mass.  Also, there is a relative deficit at low masses compared to the observed distribution that is sensitive to the amount of remaining hydrogen.  As with single stars, the BPASS distribution strongly depends on metallicity.  We show that distributions corresponding to metallicities substantially higher and lower than SLSN host galaxies provide poor matches to the SLSN distribution.          

\subsection{Angular Momentum Transport Weakens the Engine?}

In addition to the complex shape of the pre-SN mass distribution itself, we find a trend between the pre-SN masses and the magnetar initial spin periods; SLSNe with lower ejecta masses exhibit systematically slower initial spins.  This confirms initial indications of such a trend from the modeling of a smaller sample of bolometric light curves by \citet{Yu2017}.  Our analysis verifies this trend is not due to observational biases and suggests that the lowest mass pre-explosion progenitors are spun down, possibly a signature of significant mass loss.  Though the observed trend is likely the result of multiple mechanisms, including spin up processes, acting at different stages of evolution through core collapse.  

The process by which sufficient angular momentum is retained in stellar cores, in this case to form magnetars with $P\sim1-8$ ms, is not well understood.  Early models of massive single stars at solar metallicity exploring angular momentum transport effects due to magnetic torques predict initial spins of $P\gtrsim10$ ms \citep{Heger2005}, whereas more recent models including additional instabilities predict much slower spins ($P\sim50-200$ ms; \citealt{MaFuller2019}), indicating that spin up due to mass transfer or tidal effects in binary systems may play an important role \citep{HeuvelYoon2007,deMink2013}.  In addition, CHE at low metallicity may be key for the production of rapidly rotating remnants \citep{YoonLanger2005}. 

\begin{figure}[t!]
    \centering
    \includegraphics[scale=0.41]{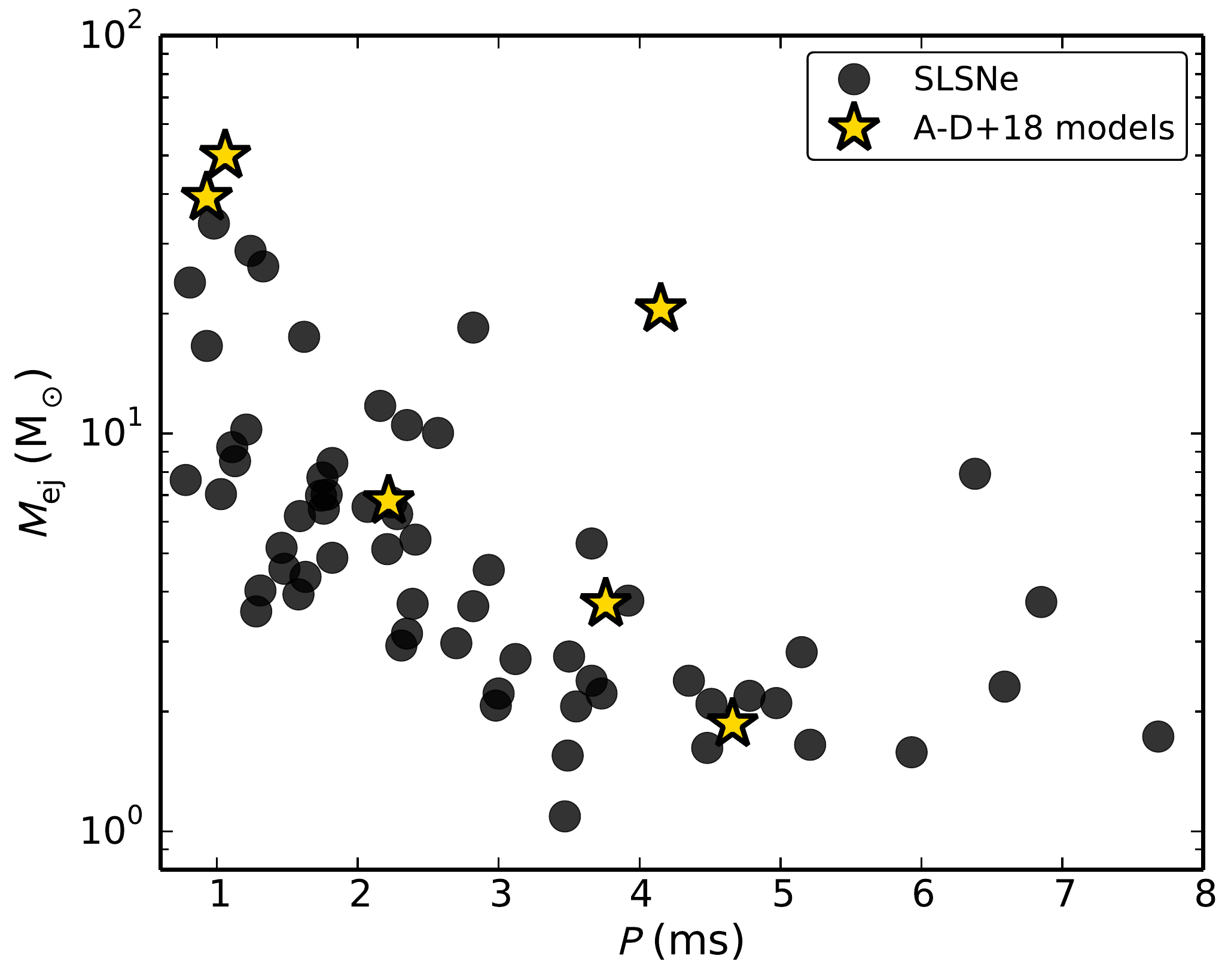}
    \caption{Comparison in $M_{\rm ej}-P$ parameter space between SLSNe (black points) and the models of \citet{AguileraDena2018} of rapidly rotating stars which leave bare CO cores due to CHE (gold stars).  In general, models with lower ejecta masses have slower spins, broadly consistent with the trend seen in the SLSN sample.}
    \label{fig:ADmodels}
\end{figure}

The \citet{AguileraDena2018} models of rapidly rotating CO cores predict a trend between mass and specific angular momentum of the core.  Specifically, the more massive models exhibit less angular momentum transport from the core to the surface, whereas lower mass models transfer significant core angular momentum leading to slower remnant spins.  In Figure~\ref{fig:ADmodels} we plot the ejecta masses and remnant spins for the \citet{AguileraDena2018} models (they assume a 1.5 M$_{\odot}$ neutron star remnant), showing this trend exhibits broad agreement with the observed SLSN mass-spin trend.

\subsection{The Black Hole Formation Threshold}

A key issue with all of the models discussed here is the threshold between neutron star and black hole formation.  Although a given progenitor evolution model may predict a final pre-SN mass distribution similar to the observed SLSN pre-explosion mass distribution, whether the further evolution through core collapse produces a black hole or a neutron star is not well understood.  Fallback accretion onto a black hole as an alternative central engine model for SLSNe \citep{DexterKasen2013} is not expected to account for the observed durations and luminosities  \citep{Moriya2018}.   Modeling by \citet{Ertl2019} of the explosions of helium stars from \citet{Woosley2019} suggest that pre-SN stars with masses above 12 M$_{\odot}$ produce black holes.  The resulting conflict between single star evolution models and the magnetar model given the observation of SLSNe with pre-SN masses extending to $\approx 40$ M$_{\odot}$, may be an indication that rapid rotation and additional effects from interactions in a binary system have a significant effect on black hole formation and the mass range where PPI becomes important. As discussed above, rotational energy may couple to the kinetic energy of the SN, enabling more massive stars to explode.  An additional possibility is that a star that undergoes PPI may later produce a magnetar-powered SLSN, a scenario possibly supported by evidence of detached CSM shells around some SLSNe \citep{Yan2017,Lunnan2018b}.  This scenario may be less and less likely as PPI effects become stronger with increasing mass, potentially reflected by the steep drop-off in the SLSN mass distribution at $M\gtrsim40$ M$_{\odot}$.

\section{Summary and Conclusions}
\label{sec:concl}

We presented an analysis of the pre-explosion progenitor mass distribution of 62 SLSNe inferred from light curve modeling in the context of a magnetar central engine.  We additionally explored possible observational biases that affect the observed mass distribution, and explored correlations between the progenitor mass and central engine properties.  Our key findings are:

\begin{itemize}

    \item The pre-explosion progenitor mass distribution spans a broad range of masses from $M\approx3.6$ M$_{\odot}$ to $M\approx40$ M$_{\odot}$, with steep drop-offs at both ends of the distribution, and evidence for a break at $\approx 9$ M$_{\odot}$.  

    \item The mass distribution is best fit with a broken power law with indices of $-0.41\pm 0.06$ at $3.6-8.6$ M$_{\odot}$ and $-1.26\pm 0.06$ at $8.6-40$ M$_{\odot}$.    
    
    \item While there is some variation in the observational efficiency of recognizing SLSNe as a function of progenitor and engine parameters, we find that this effect is mild and does not alter the broken power-law shape of the mass distribution.
    
    \item The SLSN mass distribution extends to much higher masses than that of SNe Ib/c ($\approx 40$ M$_\odot$ versus $\approx 10$ M$_\odot$, respectively), and also exhibits a different power-law slope in the overlapping mass range.  However, both distributions exhibit a steep decline at $\lesssim 3.5$ M$_\odot$. 

    \item There is a correlation between $M_{\rm ej}$ and $P$ such that events with low $M_{\rm ej}$ have systematically slower spins.  Our analysis suggests these events are spun down.
    
\end{itemize}

This analysis provides the most detailed look to date at the pre-explosion progenitor masses of SLSNe, a key property encoding information about the evolutionary history of the progenitor stars.  We further compared the pre-explosion progenitor mass distribution of SLSNe with both single and binary stellar evolution models and found the final mass distribution from models of stripped binary stars also show a break structure, whereas models of single stars do not easily explain the SLSN mass distribution.

The statistically robust break structure in the distribution and the trend between $M_{\rm ej}$ and $P$ are likely a direct result of the physical processes associated with producing a magnetar capable of enhancing the radiative output of a SN.  Non-rotating single-star models do not naturally predict such trends, whereas models of rapidly rotating stars as well as evolutionary models incorporating the effects of binary interaction may.  Our findings therefore demonstrate self-consistency of the magnetar central engine model for SLSNe.  

Independent of model comparisons, our analysis shows that the progenitors of SLSNe differ from normal stripped-envelope SNe; the effects of rotation perhaps being the key factor determining whether a stripped star explodes as a SN Ib/c, a SLSN, or at high mass, undergo direct collapse to a black hole or produce a fallback accretion powered SN.  The presence of a magnetar may naturally explain the relatively large fraction of SLSNe with massive ($M\gtrsim10$ M$_{\odot}$) pre-explosion progenitors if the conversion of rotational energy into kinetic energy enables more massive stars to explode.

\acknowledgments
P.K.B.~is supported by a CIERA Postdoctoral Fellowship.  The Berger Time-Domain Group at Harvard is supported in part by the NSF under grant AST-1714498.  M.N.~is supported by a Royal Astronomical Society Research Fellowship.  V.A.V.~is supported by the Ford Foundation through a Dissertation Fellowship.  This paper made use of the Open Supernova Catalog \citep{Guillochon2017}.

\appendix

\section{Testing for Redshift Evolution of the Mass Distribution}

\begin{figure}
    \centering
    \includegraphics[scale=0.55]{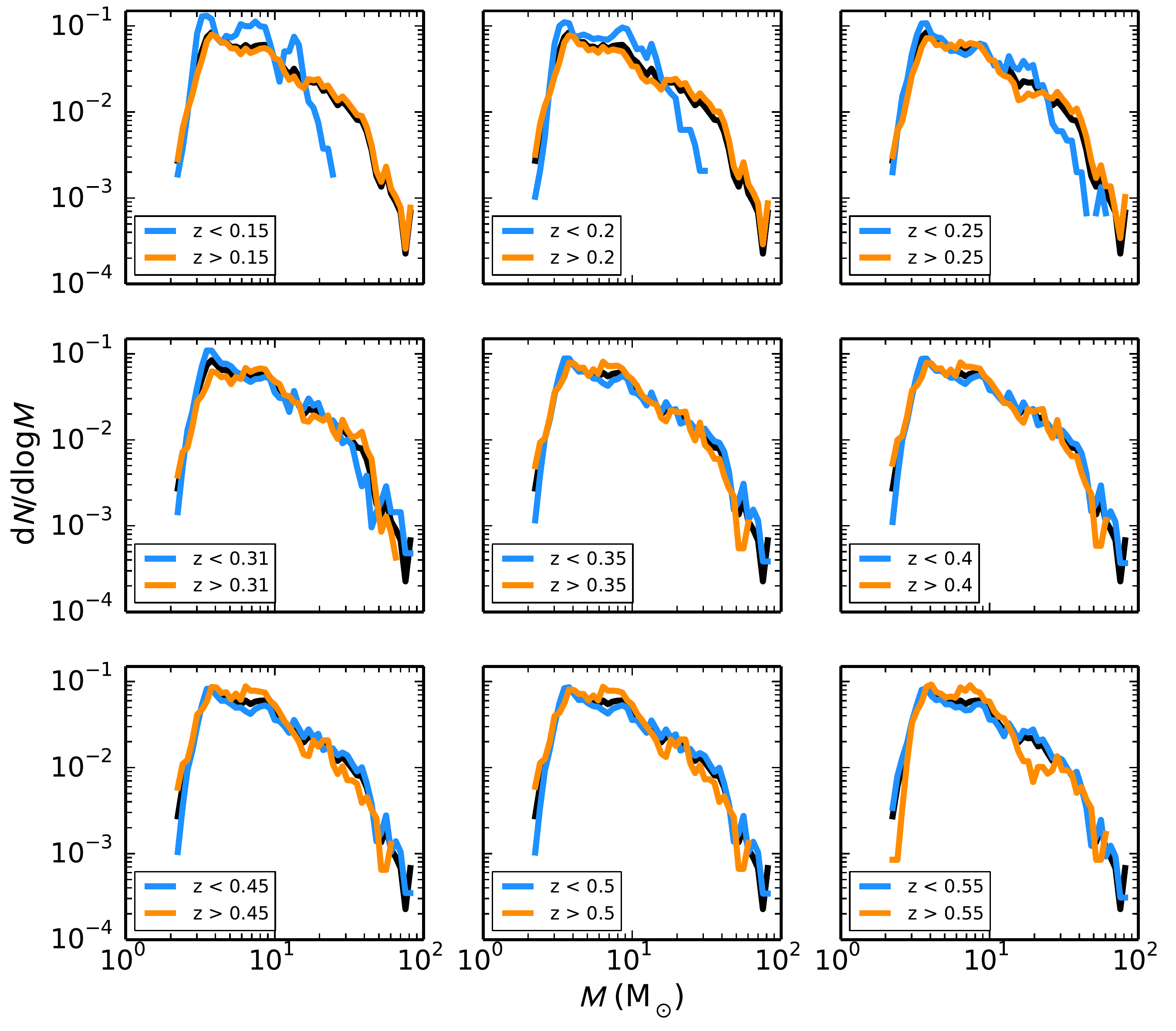}
    \caption{SLSN pre-explosion progenitor mass distribution calculated for low and high redshift sub-samples using nine different redshift cuts from $z = 0.15 - 0.55$.  The median redshift of the sample is $z\approx0.31$.  We find that dividing the sample by redshift is statistically consistent with dividing the sample randomly; in other words, we cannot discern any significant difference in the distribution at low and high redshift.}
    \label{fig:redshift}
\end{figure}

In Figure \ref{fig:redshift} we show the SLSN pre-explosion progenitor mass distribution constructed from sub-samples of SLSNe divided at various redshifts.  

\begin{thebibliography}{}

\bibitem[Aguilera-Dena et al.(2018)]{AguileraDena2018} Aguilera-Dena, D.~R., Langer, N., Moriya, T.~J., et al.\ 2018, \apj, 858, 115 

\bibitem[Angus et al.(2019)]{Angus2019} Angus, C.~R., Smith, M., Sullivan, M., et al.\ 2019, \mnras, 487, 2215

\bibitem[Barbary et al.(2009)]{Barbary2009} Barbary, K., Dawson, K.~S., Tokita, K., et al.\ 2009, \apj, 690, 1358

\bibitem[Berger et al.(2012)]{Berger2012} Berger, E., Chornock, R., Lunnan, R., et al.\ 2012, \apjl, 755, L29

\bibitem[Blanchard et al.(2018)]{Blanchard2018} Blanchard, P.~K., Nicholl, M., Berger, E., et al.\ 2018, \apj, 865, 9

\bibitem[Blanchard et al.(2019)]{Blanchard2019} Blanchard, P.~K., Nicholl, M., Berger, E., et al.\ 2019, \apj, 872, 90

\bibitem[Chatzopoulos \& Wheeler(2012)]{ChatzopoulosWheeler2012} Chatzopoulos, E., \& Wheeler, J.~C.\ 2012, \apj, 748, 42

\bibitem[Chatzopoulos et al.(2013)]{Chatzopoulos2013} Chatzopoulos, E., Wheeler, J.~C., Vinko, J., et al.\ 2013, \apj, 773, 76

\bibitem[Chen et al.(2013)]{Chen2013} Chen, T.-W., Smartt, S.~J., Bresolin, F., et al.\ 2013, \apj, 763, L28

\bibitem[Chen et al.(2015)]{Chen2015} Chen, T.-W., Smartt, S.~J., Jerkstrand, A., et al.\ 2015, \mnras, 452, 1567

\bibitem[Chen et al.(2017)]{Chen2017} Chen, T.-W., Nicholl, M., Smartt, S.~J., et al.\ 2017, \aap, 602, A9

\bibitem[Chevalier \& Irwin(2011)]{ChevalierIrwin2011} Chevalier, R.~A. \& Irwin, C.~M.\ 2011, \apj, 729, L6

\bibitem[Chomiuk et al.(2011)]{Chomiuk2011} Chomiuk, L., Chornock, R., Soderberg, A.~M., et al.\ 2011, \apj, 743, 114

\bibitem[De Cia et al.(2018)]{DeCia2018} De Cia, A., Gal-Yam, A., Rubin, A., et al.\ 2018, \apj, 860, 100

\bibitem[de Mink et al.(2013)]{deMink2013} de Mink, S.~E., Langer, N., Izzard, R.~G., et al.\ 2013, \apj, 764, 166

\bibitem[Dessart et al.(2012)]{Dessart2012} Dessart, L., Hillier, D.~J., Waldman, R., et al.\ 2012, \mnras, 426, L76

\bibitem[Dexter \& Kasen(2013)]{DexterKasen2013} Dexter, J., \& Kasen, D.\ 2013, \apj, 772, 30

\bibitem[Drout et al.(2011)]{Drout2011} Drout, M.~R., Soderberg, A.~M., Gal-Yam, A., et al.\ 2011, \apj, 741, 97

\bibitem[Eftekhari et al.(2019)]{Eftekhari2019} Eftekhari, T., Berger, E., Margalit, B., et al.\ 2019, \apjl, 876, L10

\bibitem[Eldridge et al.(2017)]{Eldridge2017} Eldridge, J.~J., Stanway, E.~R., Xiao, L., et al.\ 2017, \pasa, 34, e058

\bibitem[Ertl et al.(2019)]{Ertl2019} Ertl, T., Woosley, S.~E., Sukhbold, T., et al.\ 2019, arXiv e-prints, arXiv:1910.01641

\bibitem[Gal-Yam et al.(2009)]{Gal-Yam2009} Gal-Yam, A., Mazzali, P., Ofek, E.~O., et al.\ 2009, \nat, 462, 624

\bibitem[Gal-Yam(2012)]{Gal-Yam2012} Gal-Yam, A.\ 2012, Science, 337, 927

\bibitem[Gomez et al.(2019)]{Gomez2019} Gomez, S., Berger, E., Nicholl, M., et al.\ 2019, \apj, 881, 87

\bibitem[Greiner et al.(2015)]{Greiner2015} Greiner, J., Mazzali, P.~A., Kann, D.~A., et al.\ 2015, \nat, 523, 189

\bibitem[Guillochon et al.(2017)]{Guillochon2017} Guillochon, J., Parrent, J., Kelley, L.~Z., et al.\ 2017, \apj, 835, 64

\bibitem[Guillochon et al.(2018)]{Guillochon2018} Guillochon, J., Nicholl, M., Villar, V.~A., et al.\ 2018, The Astrophysical Journal Supplement Series, 236, 6

\bibitem[Heger \& Woosley(2002)]{HegerWoosley2002} Heger, A. \& Woosley, S.~E.\ 2002, \apj, 567, 532

\bibitem[Heger et al.(2003)]{Heger2003} Heger, A., Fryer, C.~L., Woosley, S.~E., et al.\ 2003, \apj, 591, 288

\bibitem[Heger et al.(2005)]{Heger2005} Heger, A., Woosley, S.~E., \& Spruit, H.~C.\ 2005, \apj, 626, 350

\bibitem[Howell et al.(2013)]{Howell2013} Howell, D.~A., Kasen, D., Lidman, C., et al.\ 2013, \apj, 779, 98

\bibitem[Inserra et al.(2013)]{Inserra2013} Inserra, C., Smartt, S.~J., Jerkstrand, A., et al.\ 2013, \apj, 770, 128

\bibitem[Inserra et al.(2017)]{Inserra2017} Inserra, C., Nicholl, M., Chen, T.-W., et al.\ 2017, \mnras, 468, 4642

\bibitem[Iwamoto et al.(1998)]{Iwamoto1998} Iwamoto, K., Mazzali, P.~A., Nomoto, K., et al.\ 1998, \nat, 395, 672

\bibitem[Jerkstrand et al.(2016)]{Jerkstrand2016} Jerkstrand, A., Smartt, S.~J. \& Heger, A.\ 2016, \mnras, 455, 3207

\bibitem[Jerkstrand et al.(2017)]{Jerkstrand2017} Jerkstrand, A., Smartt, S.~J., Inserra, C., et al.\ 2017, \apj, 835, 13

\bibitem[Kann et al.(2019)]{Kann2019} Kann, D.~A., Schady, P., Olivares E., F., et al.\ 2019, \aap, 624, A143

\bibitem[Kasen \& Bildsten(2010)]{KasenBildsten2010} Kasen, D., \& Bildsten, L.\ 2010, \apj, 717, 245

\bibitem[Leloudas et al.(2012)]{Leloudas2012} Leloudas, G., Chatzopoulos, E., Dilday, B., et al.\ 2012, \aap, 541, A129

\bibitem[Leloudas et al.(2015)]{Leloudas2015} Leloudas, G., Schulze, S., Kr{\"u}hler, T., et al.\ 2015, \mnras, 449, 917

\bibitem[Liu et al.(2017)]{Liu2017} Liu, Y.-Q., Modjaz, M., \& Bianco, F.~B.\ 2017, \apj, 845, 85

\bibitem[Liu et al.(2017)]{Liu2017magLCs} Liu, L.-D., Wang, S.-Q., Wang, L.-J., et al.\ 2017, \apj, 842, 26

\bibitem[Lunnan et al.(2013)]{Lunnan2013} Lunnan, R., Chornock, R., Berger, E., et al.\ 2013, \apj, 771, 97

\bibitem[Lunnan et al.(2014)]{Lunnan2014} Lunnan, R., Chornock, R., Berger, E., et al.\ 2014, \apj, 787, 138

\bibitem[Lunnan et al.(2016)]{Lunnan2016} Lunnan, R., Chornock, R., Berger, E., et al.\ 2016, \apj, 831, 144

\bibitem[Lunnan et al.(2018a)]{Lunnan2018} Lunnan, R., Chornock, R., Berger, E., et al.\ 2018a, \apj, 852, 81

\bibitem[Lunnan et al.(2018b)]{Lunnan2018b} Lunnan, R., Fransson, C., Vreeswijk, P.~M., et al.\ 2018b, Nature Astronomy, 2, 887

\bibitem[Lunnan et al.(2019)]{Lunnan2019} Lunnan, R., Yan, L., Perley, D.~A., et al.\ 2019, arXiv e-prints, arXiv:1910.02968

\bibitem[Lyman et al.(2016)]{Lyman2016} Lyman, J.~D., Bersier, D., James, P.~A., et al.\ 2016, \mnras, 457, 328

\bibitem[Ma \& Fuller(2019)]{MaFuller2019} Ma, L., \& Fuller, J.\ 2019, \mnras, 488, 4338

\bibitem[Mazzali et al.(2013)]{Mazzali2013} Mazzali, P.~A., Walker, E.~S., Pian, E., et al.\ 2013, \mnras, 432, 2463

\bibitem[Mazzali et al.(2016)]{Mazzali2016} Mazzali, P.~A., Sullivan, M., Pian, E., et al.\ 2016, \mnras, 458, 3455

\bibitem[McCrum et al.(2014)]{McCrum2014} McCrum, M., Smartt, S.~J., Kotak, R., et al.\ 2014, \mnras, 437, 656

\bibitem[McCrum et al.(2015)]{McCrum2015} McCrum, M., Smartt, S.~J., Rest, A., et al.\ 2015, \mnras, 448, 1206

\bibitem[Milisavljevic et al.(2013)]{Milisavljevic2013} Milisavljevic, D., Soderberg, A.~M., Margutti, R., et al.\ 2013, \apj, 770, L38

\bibitem[Moriya et al.(2018)]{Moriya2018} Moriya, T.~J., Nicholl, M., \& Guillochon, J.\ 2018, \apj, 867, 113

\bibitem[Nicholl et al.(2013)]{Nicholl2013} Nicholl, M., Smartt, S.~J., Jerkstrand, A., et al.\ 2013, \nat, 502, 346

\bibitem[Nicholl et al.(2014)]{Nicholl2014} Nicholl, M., Smartt, S.~J., Jerkstrand, A., et al.\ 2014, \mnras, 444, 2096

\bibitem[Nicholl et al.(2015a)]{Nicholl2015a} Nicholl, M., Smartt, S.~J., Jerkstrand, A., et al.\ 2015a, \apjl, 807, L18

\bibitem[Nicholl et al.(2015b)]{Nicholl2015} Nicholl, M., Smartt, S.~J., Jerkstrand, A., et al.\ 2015b, \mnras, 452, 3869

\bibitem[Nicholl, \& Smartt(2016)]{NichollSmartt2016} Nicholl, M., \& Smartt, S.~J.\ 2016, \mnras, 457, L79

\bibitem[Nicholl et al.(2016a)]{Nicholl2016a} Nicholl, M., Berger, E., Smartt, S.~J., et al.\ 2016a, \apj, 826, 39

\bibitem[Nicholl et al.(2016b)]{Nicholl2016b} Nicholl, M., Berger, E., Margutti, R., et al.\ 2016b, \apj, 828, L18

\bibitem[Nicholl et al.(2017a)]{Nicholl2017gaia16apd} Nicholl, M., Berger, E., Margutti, R., et al.\ 2017a, \apj, 835, L8

\bibitem[Nicholl et al.(2017b)]{Nicholl2017} Nicholl, M., Guillochon, J., \& Berger, E.\ 2017b, \apj, 850, 55

\bibitem[Nicholl et al.(2018)]{Nicholllate15bn} Nicholl, M., Blanchard, P.~K., Berger, E., et al.\ 2018, \apj, 866, L24

\bibitem[Nicholl et al.(2019)]{Nicholl2019} Nicholl, M., Berger, E., Blanchard, P.~K., et al.\ 2019, \apj, 871, 102

\bibitem[{\"O}zel \& Freire(2016)]{NSreview} {\"O}zel, F., \& Freire, P.\ 2016, \araa, 54, 401

\bibitem[Papadopoulos et al.(2015)]{Papadopoulos2015} Papadopoulos, A., D'Andrea, C.~B., Sullivan, M., et al.\ 2015, \mnras, 449, 1215

\bibitem[Pastorello et al.(2010)]{Pastorello2010} Pastorello, A., Smartt, S.~J., Botticella, M.~T., et al.\ 2010, \apj, 724, L16

\bibitem[Perley et al.(2016)]{Perley2016} Perley, D.~A., Quimby, R.~M., Yan, L., et al.\ 2016, \apj, 830, 13

\bibitem[Podsiadlowski et al.(1992)]{Podsiadlowski1992} Podsiadlowski, P., Joss, P.~C., \& Hsu, J.~J.~L.\ 1992, \apj, 391, 246

\bibitem[Prajs et al.(2017)]{Prajs2017} Prajs, S., Sullivan, M., Smith, M., et al.\ 2017, \mnras, 464, 3568

\bibitem[Prentice et al.(2019)]{Prentice2019} Prentice, S.~J., Ashall, C., James, P.~A., et al.\ 2019, \mnras, 485, 1559

\bibitem[Quimby et al.(2007)]{Quimby2007} Quimby, R.~M., Aldering, G., Wheeler, J.~C., et al.\ 2007, \apjl, 668, L99

\bibitem[Quimby et al.(2011)]{Quimby2011} Quimby, R.~M., Kulkarni, S.~R., Kasliwal, M.~M., et al.\ 2011, \nat, 474, 487

\bibitem[Quimby et al.(2018)]{Quimby2018} Quimby, R.~M., De Cia, A., Gal-Yam, A., et al.\ 2018, \apj, 855, 2

\bibitem[Salpeter(1955)]{Salpeter1955} Salpeter, E.~E.\ 1955, \apj, 121, 161

\bibitem[Sana et al.(2012)]{Sana2012} Sana, H., de Mink, S.~E., de Koter, A., et al.\ 2012, Science, 337, 444

\bibitem[Schulze et al.(2018)]{Schulze2018} Schulze, S., Kr{\"u}hler, T., Leloudas, G., et al.\ 2018, \mnras, 473, 1258

\bibitem[Smith et al.(2016)]{Smith2016} Smith, M., Sullivan, M., D'Andrea, C.~B., et al.\ 2016, \apjl, 818, L8

\bibitem[Spera et al.(2015)]{Spera2015} Spera, M., Mapelli, M., \& Bressan, A.\ 2015, \mnras, 451, 4086

\bibitem[Stanway, \& Eldridge(2018)]{StanwayEldridge2018} Stanway, E.~R., \& Eldridge, J.~J.\ 2018, \mnras, 479, 75

\bibitem[Sukhbold et al.(2016)]{Sukhbold2016} Sukhbold, T., Ertl, T., Woosley, S.~E., et al.\ 2016, \apj, 821, 38

\bibitem[Sukhbold et al.(2018)]{Sukhbold2018} Sukhbold, T., Woosley, S.~E., \& Heger, A.\ 2018, \apj, 860, 93

\bibitem[Taddia et al.(2018)]{Taddia2018} Taddia, F., Stritzinger, M.~D., Bersten, M., et al.\ 2018, \aap, 609, A136

\bibitem[van den Heuvel \& Yoon(2007)]{HeuvelYoon2007} van den Heuvel, E.~P.~J., \& Yoon, S.-C.\ 2007, \apss, 311, 177

\bibitem[Villar et al.(2018)]{Villar2018} Villar, V.~A., Nicholl, M., \& Berger, E.\ 2018, \apj, 869, 166

\bibitem[Vreeswijk et al.(2014)]{Vreeswijk2014} Vreeswijk, P.~M., Savaglio, S., Gal-Yam, A., et al.\ 2014, \apj, 797, 24

\bibitem[Vreeswijk et al.(2017)]{Vreeswijk2017} Vreeswijk, P.~M., Leloudas, G., Gal-Yam, A., et al.\ 2017, \apj, 835, 58

\bibitem[Woosley \& Heger(2006)]{WoosleyHeger2006} Woosley, S.~E., \& Heger, A.\ 2006, \apj, 637, 914

\bibitem[Woosley(2010)]{Woosley2010} Woosley, S.~E.\ 2010, \apj, 719, L204

\bibitem[Woosley(2017)]{Woosley2017} Woosley, S.~E.\ 2017, \apj, 836, 244

\bibitem[Woosley(2019)]{Woosley2019} Woosley, S.~E.\ 2019, \apj, 878, 49

\bibitem[Yan et al.(2015)]{Yan2015} Yan, L., Quimby, R., Ofek, E., et al.\ 2015, \apj, 814, 108

\bibitem[Yan et al.(2017)]{Yan2017} Yan, L., Lunnan, R., Perley, D.~A., et al.\ 2017, \apj, 848, 6

\bibitem[Yoon \& Langer(2005)]{YoonLanger2005} Yoon, S.-C., \& Langer, N.\ 2005, \aap, 443, 643

\bibitem[Yoon et al.(2006)]{Yoon2006} Yoon, S.-C., Langer, N., \& Norman, C.\ 2006, \aap, 460, 199

\bibitem[Yu et al.(2017)]{Yu2017} Yu, Y.-W., Zhu, J.-P., Li, S.-Z., et al.\ 2017, \apj, 840, 12

\end{thebibliography}
\end{document}